\documentclass[aps, reprint, pra, longbibliography]{revtex4-1}

\usepackage{hyperref}
\usepackage{graphicx}
\usepackage{amsmath}
\usepackage{wasysym}
\usepackage{amsfonts}
\usepackage{color}
\usepackage{verbatim}
\graphicspath{{./figures/}}

\newcommand{\beq}{\begin{equation}}
\newcommand{\eeq}{\end{equation}}
\newcommand{\bea}{\begin{eqnarray}}
\newcommand{\eea}{\end{eqnarray}}
\newcommand{\bal}{\begin{align}}
\newcommand{\eal}{\end{align}}

\newcommand{\eq}[1]{Eq.~(\ref{#1})}

\newcommand{\fig}[1]{Fig.~\ref{#1}}

\begin{document}

\title{Electron-phonon coupling and energy flow in a simple metal beyond the two-temperature approximation}

%\author{Lutz Waldecker}
%\email{waldecker@fhi-berlin.mpg.de}
%\author{Roman Bertoni}
%\affiliation{Fritz-Haber-Institut der Max-Planck-Gesellschaft, Faradayweg 4-6, 14195 Berlin, Germany}
%\author{Jan Vorberger}
%\affiliation{Max-Planck-Institut f\"ur Physik komplexer Systeme, N\"othnitzer Strasse 38, 01187 Dresden, Germany}
%\affiliation{Helmholtz-Zentrum Dresden-Rossendorf, Bautzner Landstrasse 400, 01328 Dresden, Germany}
%\author{Ralph Ernstorfer}
%\email{ernstorfer@fhi-berlin.mpg.de}
%\affiliation{Fritz-Haber-Institut der Max-Planck-Gesellschaft, Faradayweg 4-6, 14195 Berlin, Germany}

\author{Lutz Waldecker}
\email{waldecker@fhi-berlin.mpg.de}
\author{Roman Bertoni}
\affiliation{Fritz-Haber-Institut der Max-Planck-Gesellschaft, 14195 Berlin, Germany}
\author{Jan Vorberger}
\affiliation{Max-Planck-Institut f\"ur Physik komplexer Systeme, 01187 Dresden, Germany}
\affiliation{Helmholtz-Zentrum Dresden-Rossendorf, 01328 Dresden, Germany}
\author{Ralph Ernstorfer}
\email{ernstorfer@fhi-berlin.mpg.de}
\affiliation{Fritz-Haber-Institut der Max-Planck-Gesellschaft, 14195 Berlin, Germany}

\begin{abstract}
The electron-phonon coupling and the corresponding energy exchange was investigated experimentally and by \textit{ab initio} theory in non-equilibrium states of the free-electron metal aluminium. The temporal evolution of the atomic mean squared displacement in laser-excited thin free-standing films was monitored by femtosecond electron diffraction. The electron-phonon coupling strength was obtained for a range of electronic and lattice temperatures from density functional theory molecular dynamics (DFT-MD) simulations. The electron-phonon coupling parameter extracted from the experimental data in the framework of a two-temperature model (TTM) deviates significantly from the \textit{ab initio} values. We introduce a non-thermal lattice model (NLM) for describing non-thermal phonon distributions as a sum of thermal distributions of the three phonon branches. The contributions of individual phonon branches to the electron-phonon coupling are considered independently and found to be dominated by longitudinal acoustic phonons. Using all material parameters from first-principle calculations besides the phonon-phonon coupling strength, the prediction of the energy transfer from electrons to phonons by the NLM is in excellent agreement with time-resolved diffraction data. Our results suggest that the TTM is insufficient for describing the microscopic energy flow even for simple metals like aluminium and that the determination of the electron-phonon coupling constant from time-resolved experiments by means of the TTM leads to incorrect values. In contrast, the NLM describing transient phonon populations by three parameters appears to be a sufficient model for  quantitatively describing electron-lattice equilibration in aluminium. We discuss the general applicability of the NLM and provide a criterion for the suitability of the two-temperature approximation for other metals.
\end{abstract}

\maketitle
\date{\today}

\section{Introduction}
The interaction between electrons and lattice vibrations is central to both ground state as well as out-of-equilibrium properties of solids. The huge phase space of elementary energy excitations of electronic and phononic origin renders a universally valid description of electron and phonon distributions virtually impossible. The most common approach is to assume thermal Fermi-Dirac and Bose-Einstein distributions for electrons and phonons, respectively, and to describe the non-equilibrium quantum statistics by a Boltzmann equation (see e.g. \cite{book:Grimvall, book:Patterson}). These approximations are employed in the description of near-equilibrium phenomena like electrical or thermal conductivity as well as in the treatment of states with pronounced non-equilibrium distributions.

Non-equilibrium states occur after impulsive perturbation of a material, for instance, after excitation with femtosecond laser pulses. For the case of metals, Anisimov et al.~introduced an empirical two-temperature model (TTM) describing the energy transfer from the excited electrons to the lattice with a single electron-phonon coupling parameter~\cite{Anisimov1967, Anisimov1974}. Allen refined this approach by developing a microscopic theory and derived the electron-phonon coupling parameter as the spectral integral of the zero-temperature Eliashberg spectral function $\alpha^2(\omega) F(\omega)$~\cite{Allen1987}. $\alpha^2F(\omega)$ comprises the phonon density of states $F(\omega)$ and the phonon frequency-dependent electron-phonon coupling $\alpha^2(\omega)$, which incorporates all allowed scattering processes of electrons with phonons of frequency $\omega$. While Allen's work provides a link between the Eliashberg theory of conventional superconductors and the electron-lattice equilibration in laser-excited metals, the underlying assumptions, in particular the independence of $\alpha^2$ on the electron temperature  and the assumption of thermal distributions of electrons and phonons at all times, are questionable in the case of pronounced non-equilibrium situations \cite{Lin2008, Rethfeld2002}. 
%%%%%%%%%%%%%%%% PhDOS-a2F %%%%%%%%%%%%%%%%%%%%%%%%%%%%%%%%%%%%%%%%%%%%%%%%%%%%%%%%%%%
\begin{figure}[htb]
\begin{center}
% reprint
\includegraphics[width=0.9\columnwidth]{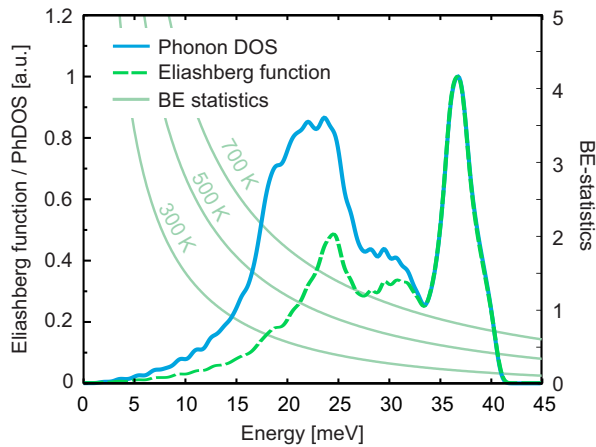}
\caption{Normalized phonon density of states (PhDOS) and Eliashberg function of aluminium, calculated with DFT and the \textit{abinit} package. The solid green lines show Bose-Einstein statistics for three lattice temperatures.}
\label{fig:motivation}
\end{center}
\end{figure}
%%%%%%%%%%%%%%%%%%%%%%%%%%%%%%%%%%%%%%%%%%%%%%%%%%%%%%%%%%%
This is rationalized by comparing the phonon density of states (PhDOS) and the Eliashberg function, see the results of our DFT calculation in \fig{fig:motivation}: in a thermal state, phonons follow Bose-Einstein (BE) statistics (thin lines, plotted for three temperatures), and the phonon occupation is given by the phonon DOS multiplied with the BE statistics. The coupling of electrons to phonons is given by the Eliashberg function and is strongest for the high-energy phonons, whose occupation is the lowest in a thermal state. This implies that energy transfer between electrons and lattice in non-equilibrium states leads to transient non-thermal phonon distributions.

In the past, a range of experimental techniques have been employed to investigate the electron-phonon coupling of metals quantitatively in equilibrium and non-equilibrium situations. Among the first were measurements of the phonon lifetime by neutron diffraction~\cite{Stedman1967}, which, however, depend on electron-phonon and phonon-phonon scattering rates. The Eliashberg functions of metals have been determined by electron tunneling spectroscopy~\cite{McMillan1965}, which have to be done in the superconducting state at low temperatures. Angle-resolved photoelectron spectroscopy has been used to measure the electron-phonon coupling near surfaces through the experimental determination of the electron self-energy~\cite{Plummer2003}.

The electron-phonon coupling in non-equilibrium situations has been studied by time-resolved measurements. Subsequent to impulsive excitation of electrons by a femtosecond laser pulse, the electron-phonon coupling governs the dynamics of equilibration of electrons and lattice. The most common approach is time-domain thermoreflectance measurements \cite{Elsayed-Ali1987, Schoenlein1987, Brorson1990}. While these measurements are easy to implement, the relation between spectroscopic observables, e.g.~the reflectivity, and physical quantities of the non-equilibrium system is, in general, nontrivial. In particular, the respective contributions of electronic excitation, increasing phonon population and thermal expansion to the transient reflectivity are \textit{per se} unknown and approximated by linear dependencies. Time-resolved photoelectron spectroscopy provides a more direct access to transient electron distributions which can be related to the electron-phonon interaction~\cite{Sentef2013}. In the case of simple metals, however, its intrinsic surface sensitivity makes a disentanglement of electron relaxation due to coupling phenomena and ballistic electron transport out of the probed sample volume virtually impossible~\cite{Bauer1998,Hopkins2009}. In contrast, time-resolved diffraction provides direct information on the vibrational excitation and long-range order of the crystalline structure. In fact, the first studies of photo-induced phase transitions with time-resolved electron diffraction have been performed on thin aluminium samples~\cite{Williamson1984, Siwick2003}. However, the quantitative investigation of electron-phonon coupling in simple metals requires a high temporal resolution not available in time-resolved electron diffraction until recently.

In this article, we report a combined experimental and theoretical study of the energy transfer from photo-excited electrons to the lattice in the quasi-free-electron metal aluminium by femtosecond electron diffraction (FED) and density functional theory (DFT). The experimental and theoretical approaches are described in sections~\ref{sec:exp} and~\ref{sec:th}. In section~\ref{sec:TTM}, we first apply a two-temperature model to extract electron-phonon coupling constants from the diffraction data and find a significant deviation from the values obtained from \textit{ab initio} calculations. In section~\ref{sec:NLM}, we introduce a non-thermal lattice model {allowing for an approximate description of} non-equilibrium phonon distributions. The predictions of this model are compared to the experimentally observed atomic mean squared displacements (MSD) and a quantitative agreement of DFT calculations and measurements is found.

\section{Experimental methods}
\label{sec:exp}
\begin{figure}[bth]
\begin{center}
% reprint
\includegraphics[width=1.0\columnwidth]{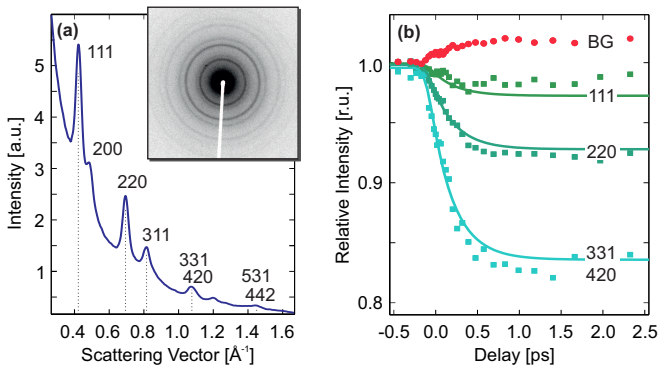}
\caption{(a) Diffraction image of 30\,nm thick polycristalline aluminium (inset) and radial average, retrieved by angular integration. Peaks with multiple labels are too close together to separate experimentally. (b) Evolution of the relative intensity of three selected Bragg peaks and the background (BG) of diffusely scattered electrons with respect to negative time delays for an absorbed energy density of 440\,J/cm$^3$. The solid lines represent a global fit to the data (see text for details).}
\label{fig:Pdecay}
\end{center}
\end{figure}

The femtosecond electron diffraction measurements were conducted with a compact electron diffraction setup described in detail elsewhere~\cite{Waldecker2015Setup}. We use near-infrared pump pulses centered at 800\,nm with a duration of approximately 50\,fs (FWHM) to photoexcite 30\,nm thick films of free-standing, polycrystalline aluminium (Plano GmbH) and femtosecond electron pulses to probe their atomic structure at different delay times. The pump spot size is 250\,$\mu$m FWHM and about twice as big as the probe spot. The energy of the pump pulses was varied to achieve absorbed energy densities between 125\,J/cm$^3$ and 840\,J/cm$^3$. The repetition rate of the laser was reduced from 1 to 0.5\,kHz for the highest fluences to allow for thermal relaxation back to room temperature between laser pulses. The probe electrons have a kinetic energy of 100\,keV and the pulses contain a few-thousand electrons. According to numerical simulations, the electron pulse duration at the sample position is approximately 100\,fs~\cite{Waldecker2015Setup}. The diffracted electrons are detected with an electron camera and the images are angularly integrated to retrieve radial averages, shown in \fig{fig:Pdecay}(a) (diffraction image in the inset). {Comparison to powder x-ray diffraction data reveals that the polycrystalline films exhibit partial texture with preferred (100)-direction.}

%%%%%%%%%%%%%%%% Peak height %%%%%%%%%%%%%%%%%%%%%%%%%%%%%%%%%%%%%%%%%%%%%%%%%%%%%%%%%%%
%%%%%%%%%%%%%%%%%%%%%%%%%%%%%%%%%%%%%%%%%%%%%%%%%%%%%%%%%%%

For each pump-probe delay, we fit an empirical background function (Lorentzian + fourth order polynomial) and pseudo-voigt line-profiles to the peaks in the radial averages. The fitting of background and Bragg peaks is refined iteratively. The temporal evolution of the relative intensity of three selected Bragg peaks, as well as changes in the integrated background between scattering vector $s=0.2$\AA$^{-1}$ and 1.95\,\AA$^{-1}$, are plotted as a function of pump-probe delay in \fig{fig:Pdecay}(b) for an absorbed energy density of 440\,J/cm$^3$. The intensity of all peaks decreases after photoexcitation at $t=t_0$ and reaches a new steady-state in less than one picosecond. On the same timescale, the number of diffusely scattered electrons increases, resulting in a dynamic background signal. \\
The intensity of scattered electrons into a Bragg diffraction peak depends on the atoms' movements, i.e.~their MSD $\langle u^2 \rangle$ from their equilibrium position (see e.g. Ref.~\cite{book:Peng})
\begin{equation}
\label{eq:DWF}
I = I_0 \cdot \exp \left ( -\frac43 \pi^2 \langle u^2 \rangle s^2 \right ),
\end{equation}
where $I_0$ is the scattered intensity of a perfect lattice and $s = 2\sin{(\Theta)}/\lambda$ denotes the scattering vector. The observed decrease in intensity can thus be attributed to an energy transfer from excited electrons to the lattice, leading to a higher MSD of the atoms. \\
The measured relative intensities of the Bragg peaks $I_{rel}(t) = I(t)/I(t<t_0)$ can be represented in a form that does not depend on the scattering vector and therefore allows for averaging of all fitted peak heights
\begin{equation}
\label{eq:BTt}
- \frac{3}{4\pi^2} \frac{\ln (I_{\mathrm{rel},s}(t))}{s^2} = \langle u^2 \rangle (t) - \langle u^2 \rangle (t < t_0) .
\end{equation}
We include the (111), (220) and (311) reflections as well as the averaged (331)/(420) and (531)/(442) reflections, which are too close together to be separated in our experiment, in the data analysis. {As the analysis only relies on relative changes of diffraction intensities, the partial texture of the samples does not affect the time-dependent signal.}
We fit {the quantity of Eq.}~(\ref{eq:BTt}) with a mono-exponential function, convoluted with a Gaussian function of 150\,fs FWHM to account for the system response. This yields a time constant of $\tau = 350\pm45$\,fs for the dataset shown. The background rises with a time constant of $\tau = 270\pm20$\,fs. We note that these time constants are significantly shorter than previously reported values for the lattice heating time constant of ($550\pm80$\,fs,~\cite{Nie2006}), the Bragg peak decay time of aluminum ($1.0\pm0.1$\,ps,~\cite{Zhu2013}) and the rise of the inelastically scattered intensity ($0.7\pm0.1$\,ps,~\cite{Zhu2013}) at comparable excitation levels. The solid lines in \fig{fig:Pdecay}(b) are derived from the global fit by inversion of \eq{eq:BTt} and reproduce the behaviour of each diffraction peak. 
%\add{The fact that the amplitudes are well described by the global fit suggests that the observed lattice vibrations are isotropic.}

\section{Theory}
\label{sec:th}

The time evolution of a system of electrons and phonons including electron-phonon coupling can be studied theoretically by means of a set of Bloch-Peierls-Boltzmann equations as proposed by Allen \cite{Allen1987}. This system of equations conserves the energy that is transferred between electrons and phonons. In the spirit of the two-temperature model, the change in temperature is determined by the energy transfer from electrons to the lattice $Z_{ep}$ and vice versa.
The electron-phonon energy transfer rate is then given by a moment of the Bose distribution function $n_B(q,t)$ for the phonons \cite{Allen1987}
\beq \label{zeph}
Z_{ep}(T_e,T_l,t)=\frac{\partial E_{ph}(t)}{\partial t}=\sum_q
\hbar\omega(q,T_e,T_l,t)\frac{\partial n_B(q,t)}{\partial t}\,.
\eeq
Note that the phonon energies $\hbar\omega$ and therefore the energy transfer rate depend on the temperatures of the system. $Z_{ep}$ can be obtained from the appropriate Boltzmann equation and reads in terms of the electron-phonon matrix element $M_{kk'}^q$ \cite{Allen1987}
\bea
Z_{ep}(T_e,T_l,t)&=&\frac{4\pi}{\hbar}\sum_{qk}\hbar\omega_q(T_e,T_l,t)|M_{kk'}^q(t)|^2S(k,k',t)\nonumber\\
&&\times\delta\big(\varepsilon(k)-\varepsilon(k')+\hbar\omega(q,T_e,T_l,t)\big).
\eea
The phonon and electron wave vectors are connected via $k-k'=q$ and the difference in single-electron energies $\varepsilon(k)$ has to match the phonon energy $\hbar\omega(q)$. The thermal factor $S(k,k')$ is given by
\bea
S(k,k',t)&=&\left[f(k,t)-f(k',t)\right]n_B(q,t)\nonumber\\
&&-f(k',t)\left[1-f(k,t)\right]
\eea
and accounts for Pauli blocking in the scattering process of electrons via the Fermi functions $f(k,t)$.
The standard approach in calculating this expression is to introduce the Eliashberg function as a generalized electron-phonon matrix element \cite{Allen1987}
\bea
\lefteqn{\alpha^2F(\varepsilon,\varepsilon',\omega)=\frac{2}{\hbar N_c^2g(\varepsilon_F)}}&&\nonumber\\
&&\times \sum_{kk'}
|M_{kk'}^q|^2\delta(\omega-\omega(q))\delta(\varepsilon-\varepsilon(k))\delta(\varepsilon'-\varepsilon(k'))\,.
\eea
Here, $g(\varepsilon_F)$ denotes the value of the electronic density of states (DOS) at the Fermi edge.
We observe that the Eliashberg function varies on a meV scale compared to the usual eV scale for the electronic DOS. Following Wang {\em et al.} \cite{Wang:1994}, we therefore introduce the first approximation
\bea
\alpha^2F(\varepsilon,\varepsilon',\omega)&=&
\frac{g(\varepsilon)g(\varepsilon')}{g^2(\varepsilon_F)}\alpha^2F(\varepsilon_F,\varepsilon_F,\omega)\nonumber\\
&=&\frac{g(\varepsilon)g(\varepsilon')}{g^2(\varepsilon_F)}\alpha^2F(\omega)\,.
\eea
We also use, based on the same argument of scale between the phonon energies and the electronic energies
\beq
f_e(\varepsilon-\hbar\omega)-f_e(\varepsilon)=-\hbar\omega\frac{\partial f_e(\varepsilon)}{\partial \varepsilon}\,,
\eeq
and $g(\varepsilon_F+\hbar\omega(q))\sim g(\varepsilon_F)$.
This results in \cite{Lin2008}
\bea
\lefteqn{Z_{ep}(T_e,T_l,t)=-\frac{2\pi N_c}{g(\varepsilon_F)}}&&\nonumber\\
&&\times\int\limits_0^{\infty}d\omega (\hbar\omega)^2
\alpha^2F(\omega,T_e,T_l,t)\Big[n_B^e(\omega,T_e)-n_B^{p}(\omega,T_l)\Big]\nonumber\\
&&\times\int\limits_{-\infty}^{\infty}d\varepsilon\, g^2(\varepsilon)
\frac{\partial f_e(\varepsilon,T_e)}{\partial \varepsilon}\,.
\label{zeph_full}
\eea
The Bose functions contain the electron temperature in $n_B^e$ and the lattice temperature in $n_B^{p}$. Input quantities are therefore the electron DOS $g(\varepsilon)$ and the Eliashberg function $\alpha^2F(\omega)$. 

The heat capacities can be obtained from the various density of states. For the electrons, one finds
\beq
C_e(T_e,t)=\int\limits_{-\infty}^{\infty}\frac{\partial f(\varepsilon,T_e,t)}{\partial T_e}
g(\varepsilon)\varepsilon d\varepsilon\,.
\eeq
Similarly, the lattice heat capacity is given by
%\beq
%C_l(T_l,t)=\int\limits_{-\infty}^{\infty}\frac{\partial n_B(\varepsilon,T_l,t)}{\partial T_l}
%h(\varepsilon)\varepsilon d\varepsilon\,,
%\eeq
\beq
C_l(T_l,t)=\int\limits_{-\infty}^{\infty}\frac{\partial n_B(\hbar \omega,T_l,t)}{\partial T_l}
F(\omega)\hbar \omega d\omega\,,
\eeq

%\beq
%C_p(T_p,t)=\int\limits_{-\infty}^{\infty}\frac{\partial n_B(\varepsilon,T_p,t)}{\partial T_p}
%h(\varepsilon)\varepsilon d\varepsilon\,,
%\eeq
where we denote the phonon DOS with $F(\omega)$.

The calculated energy-transfer rates $Z_{ep}$ allow to connect the microscopic theory to the diffraction experiments. The standard approach in describing the temporal evolution of the excited material is the two-temperature model, which assumes all phonons to follow Bose statistics at all times. In the following sections, we compare the TTM to our data and then expand the model to investigate the implications of a non-thermal lattice on the temporal evolution of electron and phonon distributions. We assume the electronic subsystem to be thermal at all times in both models, since the electronic potential does not significantly change for excitations up to 6\,eV \cite{Recoules2006} and the non-equilibrium electron distribution induced by the laser pulses relaxes on the timescale of 10\,fs~\cite{Bauer1998, Mueller2013}, which is much shorter than the timescales of interest of this work. Additionally, we assume that the electronic excitation equilibrates spatially on the sub-100\,fs time scale due to fast electron transport~\cite{Bauer1998} in the surface normal direction.

\section{Two-temperature model}
\label{sec:TTM}

%%%%%%%%%%%%%%%% TTM %%%%%%%%%%%%%%%%%%%%%%%%%%%%%%%%%%%%%%%%%%%%%%%%%%%%%%%%%%%
\begin{figure*}[*tbh]
\begin{center}
\includegraphics[width=1.0\textwidth]{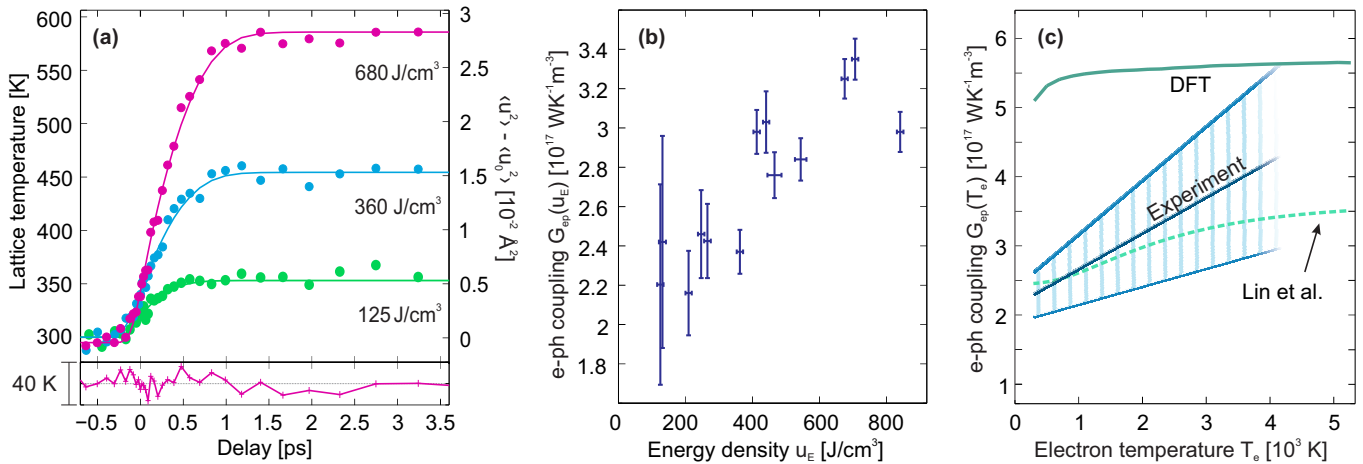}
\caption{(a) Measured lattice temperature (circles) and simulated lattice temperature with a TTM with optimized parameters for three different absorbed energy densities $u_E$. In the bottom, the residual of the TTM and the data of the highest excitation density is shown. (b) Electron-phonon coupling parameter $G_{ep}$ as a function of absorbed fluence. (c) Electron-phonon coupling dependence on electronic temperature obtained by our measurement. The green line shows $G_{ep}(T_e)$ obtained from DFT calculations by evaluating \eq{zeph_full}. The dashed light green line is the result from Lin \textit{et al.} \cite{Lin2008}.}
\label{fig:TTM}
\end{center}
\end{figure*}
%%%%%%%%%%%%%%%%%%%%%%%%%%%%%%%%%%%%%%%%%%%%%%%%%%%%%%%%%%%

In the TTM, the temporal evolutions of the thermal electron and phonon distributions are described by two coupled differential equations with the energy flow between the subsystems being proportional to the electron-phonon coupling parameter $G_{ep} = Z_{ep}/(T_e-T_l)$ and the temperature difference $T_e(t)-T_l(t)$: %and the heat-capacities $C_e$ and $C_l$
\begin{subequations} \label{eq:TTM}
\begin{align}
C_e(T_e) \frac{\partial T_e}{\partial t} &= G_{ep} \cdot (T_e-T_l) + f(t-t_0)\,, \\
C_l(T_l) \frac{\partial T_l}{\partial t} &= G_{ep} \cdot (T_e-T_l)\,.
\end{align}
\end{subequations}
The function $f(t-t_0)$ models the energy input from the laser to the electrons. Its integral is the absorbed energy density $u_E$ and its temporal shape is modeled by a gaussian function of 50 fs FWHM centered at $t=t_0$.

The predictions of the TTM can be compared to the time-resolved diffraction data by extracting the atomic MSD from the measured Bragg-peak intensities according to \eq{eq:BTt}, and the subsequent conversion to lattice temperature using a fourth-order polynomial parametrization of the (equilibrium) Debye-Waller B-factor \cite{book:Peng}. A modification of the B-factor due to phonon renormalization caused by electronic excitation is expected to be marginal as the phonon band structure is independent on electronic excitation in aluminium~\cite{Recoules2006}. The lattice temperatures reached at long delay times are in good agreement with the expected temperature rise calculated from the laser fluence and the static heat capacity of aluminium.

We continue by solving the differential equations (\ref{eq:TTM}) numerically to obtain $T_e(t)$ and $T_l(t)$ with $C_e$ and $C_l$ taken from the DFT calculations as described in section~\ref{sec:th}. With $G_{ep} = G_{ep}(u_E)$, $u_E$ and the zero time of the experiment as free parameters, we minimize the difference in $T_l(t)$ between experiment and TTM. Three data sets are exemplified in \fig{fig:TTM}(a) together with the best fits of the TTM. For all employed excitation densities, we find that the TTM is able to excellently reproduce the time-dependent data using the fitted coupling parameters shown in \fig{fig:TTM}(b).
%Using the electron-phonon coupling from the DFT calculations as an input to the TTM, however, we find that the model predicts a rise time of the lattice temperature which is too short in comparison to the data.
The increase of $G_{ep}(u_E)$ with $u_E$ suggests a significant dependence of $G_{ep}$ on $T_e$, as introduced by Lin \textit{et al.}~\cite{Lin2008}, since larger $u_E$ results in transiently larger $T_e$. We thus include a $T_e$-dependence of the coupling parameter, $G_{ep} = G_{ep}(T_e)$, in the TTM. We restrict the dependence to being linear, $G_{ep}(T_e) = g_0 + g_1\cdot T_e$, in order to keep the complexity of the fit to a manageable level. The result of global fitting of all data sets is shown in \fig{fig:TTM}(c) as solid blue line with the shaded area indicating the standard deviation. The data is plotted up to $T_e=4000$~K, which corresponds to the highest $T_e(t)$ reached transiently in the measurements.
While the experimentally determined $G_{ep}(T_e)$ is in very good agreement with the theoretical results from Ref.~\cite{Lin2008}, we observe qualitative and quantitative disagreement with $G_{ep}(T_e)$ calculated in this work (green solid line in \fig{fig:TTM}(c)): i) the low-excitation limit of the experimental $G_{ep}$ is approximately a factor of two smaller than the value obtained from \textit{ab initio} calculations; ii) the $T_e$-dependence of the theoretical $G_{ep}(T_e)$ is much less pronounced compared to the experimental results.

We consider the good agreement between our experimental results and the theoretical results of Ref.~\cite{Lin2008} to be artificial, as we discuss in detail in the appendix. In brief, the function $G_{ep}(T_e)$ reported in~\cite{Lin2008} does not purely result from first-principles calculations but employs an input value $G_{ep}(T_e=300K)$ obtained from thermoreflectance measurements {employing} a TTM analysis~\cite{Hostetler1999}. The agreement in the low-$T_e$ onset of $G_{ep}$ compared to Ref.~\cite{Lin2008} {therefore does not proof the validity of the TTM but only indicates consistency of our data with the experimental results of} Ref.~\cite{Hostetler1999}. Additionally, we asign the exact shape of the $T_e$-dependence of $G_{ep}$ reported in Ref.~\cite{Lin2008} to an insufficient number of $k$-points resulting in differences in the computed electronic structure of aluminium.

Summarizing this section, we observe a significant disagreement between the $G_{ep}(T_e)$ extracted from time-resolved diffraction data in the framework of a TTM and the respective function obtained from first principles calculations.
In the following section, we introduce an alternative approach that lifts the assumption of incessant thermal distribution of the phonons.

\section{Non-thermal lattice model}
\label{sec:NLM}

%%%%%%%%%%%%%%%% theory %%%%%%%%%%%%%%%%%%%%%%%%%%%%%%%%%%%%%%%%%%%%%%%%%%%%%%%%%%%
\begin{figure*}[*tbh]
\begin{center}
\includegraphics[width=1.0\textwidth]{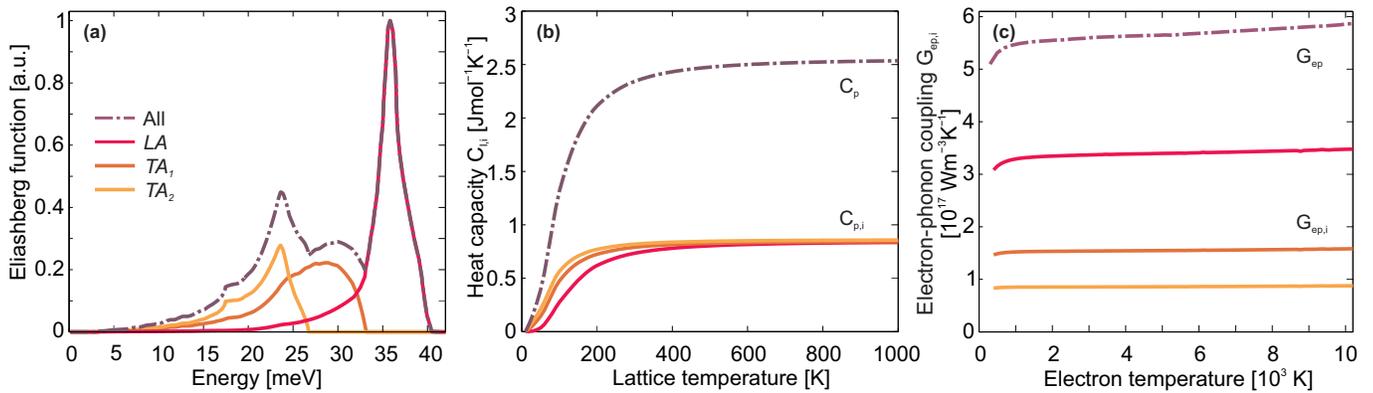}
\caption{(a) DFT calculation of the Eliashberg function and its contributions from three phonon branches $\alpha^2F_i(\omega)$, with $i=T_1,T_2,L$. (b) Lattice heat capacities of the three branches and of the entire lattice in equilibrium as a function of temperature. (c) Electron-phonon coupling calculated for the entire lattice and the individual phonon branches as function of $T_e$.}
\label{fig:th}
\end{center}
\end{figure*}
%%%%%%%%%%%%%%%%%%%%%%%%%%%%%%%%%%%%%%%%%%%%%%%%%%%%%%%%%%%

The comparison of Eliashberg function and phonon DOS shown in \fig{fig:motivation} reveals an enhanced coupling to phonons of high frequencies. This implies that laser-induced non-equilibrium between electrons and phonons subsequently leads to non-thermal phonon distributions, i.e., transiently, the phonon distribution is not resembled by the product of a Bose-Einstein function and the phonon DOS. {It is well established that non-thermal phonon populations occur after photo-excitation of materials with strong coupling of electrons to specific optical phonons. Examples include graphite exhibiting strongly coupled optical phonons}~\cite{Schafer2011, Chatelain2014} {and crystals susceptible to displacive excitation of coherent phonons (DECP) like Peierls-distorted crystals}~\cite{KST2003, Zijlstra2006} {and certain transition metal oxides}~\cite{Bothschafter2013}. {The energy flow between electrons and vibrational degrees have been described by modifications of the TTM. For the case of Peierl-dostorted lattices, the TTM has been expanded by an explicit description of the optical phonon mode linked to DECP}~\cite{Giret2011, Arnaud2013}.

In the following, we introduce a minimal model beyond the two-temperature approximation allowing for an approximate description of non-thermal phonon distributions in a simple metal where no optical phonons are present. For three-dimensional crystals with monoatomic unit cells, the phonons uniquely group into three acoustic branches with distinct dispersion relations purely due to symmetry. The partial contributions of the branches to the phonon DOS have been studied by static diffraction techniques and have been found to be well separated in frequency~\cite{Walker1956, Stedman1967}. Similarly, the contribution of the three phonon branches to the Eliashberg function can be identified. Dacorogna \textit{et al.}~calculated the \textbf{q}- and phonon branch-resolved electron-phonon coupling in aluminium along high-symmetry directions and conjectured that there is no dominant contribution from a specific phonon branch~\cite{Dacorogna1985}. Our DFT calculations performed in the full Brillouin zone{, however, reveal a pronounced phonon branch-dependence of $G_{ep}$}. Figure \ref{fig:th}(a) shows the individual contributions {$\alpha^2F_i(\omega)$} of the three phonon branches to the Eliashberg function, calculated as described in the appendix.
There is a clear spectral separation of the three contributions, in particular, the high frequency peak dominating $\alpha^2 F$ can completely be attributed to the longitudinal acoustic phonon branch (\textit{LA}).
We use \eq{zeph_full} with the partial Eliashberg functions {$\alpha^2F_i(\omega)$} to calculate the heat capacities $C_{p,i}$ and coupling parameters $G_{ep, i}$ of the three branches $i=$\textit{TA$_1$,TA$_2$,LA} as a function of $T_e$ and compare these to the values of the entire phonon spectrum in Figs.~\ref{fig:th}(b)\,\&\,(c). Whereas the partial heat capacities are very similar for all $T_l$ relevant in this work, the electron-phonon coupling differs by up to a factor four between the\textit{TA$_2$}\ and \textit{LA} branch. 

%%%%%%%%%%%%%%%% Non-thermal %%%%%%%%%%%%%%%%%%%%%%%%%%%%%%%%%%%%%%%%%%%%%%%%%%%%%%%%%%%
\begin{figure}[tbh]
\begin{center}
\includegraphics[width=1.0\columnwidth]{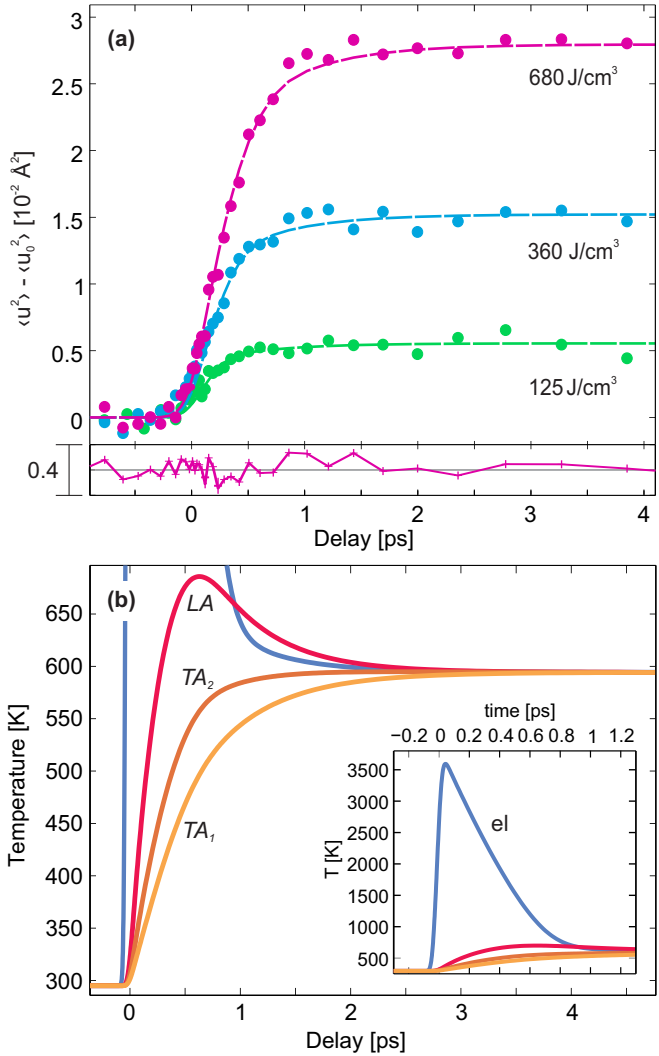}
\caption{(a) Measured mean square displacement (MSD) as a function of delay (circles) and calculated evolution predicted by the non-thermal lattice model with a phonon-phonon coupling of $3.5\cdot10^{17}$\,Wm$^{-3}$K$^{-1}$ (lines). All other material parameters were taken from DFT calculations. In the bottom, the residuals for the highest excitation density are shown. (c) Evolution of the temperatures of electrons (blue) and the three phonon branches according to the NLM for an excitation density of 680~J/cm$^3$. The inset shows the full range of $T_e(t)$. }
\label{fig:4tm}
\end{center}
\end{figure}
%%%%%%%%%%%%%%%%%%%%%%%%%%%%%%%%%%%%%%%%%%%%%%%%%%%%%%%%%%%

Consequently, we employ the natural grouping of the phonons into its three acoustic branches and introduce a non-thermal lattice model (NLM) that accounts for non-equilibrium between them. Each phonon subset is assumed to be thermal and defined by the temperature $T_{p,i}$. {This is justified by the fact that the spectral shape of the $\alpha^2 F_i(\omega)$ agrees significantly better with the respective partial phonon density of states $F_i(\omega)$ compared to the overall functions $\alpha^2F$ and $F$, as shown in Fig.}~\ref{fig:partial} {in the appendix. Additionally, for a given temperature, the Bose-Einstein function varies less over the spectral width of any $F_i$ compared to the full phonon spectrum.} The NLM therefore is a refinement of the TTM. We use \eq{zeph_full} to calculate the energy transfer to subsets of phonons by replacing the Eliashberg function $\alpha^2F(\omega)$ with a reduced counterpart, $\alpha^2F_i(\omega)$, only containing the coupling of electrons to the respective phonon branch. The energy flow is determined by their temperature difference and the evolution of all subsystems is given by the set of coupled differential equations
\begin{subequations}
\begin{align}
\label{eq:NLM}
C_e(T_e) \frac{\partial T_e}{\partial t} =& \sum_{i=1}^N G_{ep, i}(T_e) \cdot (T_e-T_{p, i}) \nonumber \\
& + f(t-t_0), \\
C_{p,i}(T_{p,i}) \frac{\partial T_{p,i}}{\partial t} =&\,G_{ep, i}(T_e) \cdot (T_e-T_{p,i}) \nonumber \\
& + \sum_{j \neq i} G_{pp, ij} \cdot (T_{p,j}-T_{p,i}).
\end{align}
\end{subequations}
where $G_{ep,i}$ is the electron-phonon coupling parameter of the $i$-th phonon subsystem to the electrons. The total energy transfer between electrons and lattice is then simply given by the sum of all partial energy transfers.
We introduced a phonon-phonon coupling parameter $G_{pp, ij}$ to account for phonon-phonon scattering processes, leading to a direct energy-transfer between branches $i$ and $j$. Due to a lack of experimental or theoretical results, we choose to use the same phonon-phonon coupling between all branches $G_{pp, ij} = G_{pp}$, the magnitude of which we determine by fitting to the experimental data. {Note that, in the limit of an infinitely large $G_{pp}$, the NLM converges to the TTM.}

In the situation of a non-thermal lattice, a conversion of the measured MSD to a lattice temperature via the Debye-Waller B-factor is invalid. Instead, the contributions of each phonon branch to the MSD can be calculated from the respective temperature calculated by the NLM. The connection of temperature and MSD is given by \cite{Sears1991}:
\begin{equation}
\langle u^2 \rangle = \frac{3\hbar}{2M} \int \limits_{0}^{\infty} \coth{\left ( \frac{\hbar \omega}{2k_BT} \right )} \frac{F \left (\omega \right )}{\omega} \, d\omega,
\end{equation}
where $M$ is the mass of an atom. By integrating over the partial DOS $F_i(\omega)$ and using the respective temperature $T_{p,i}$, we obtain the contribution of all three phonon branches $\langle u^2_i \rangle$ to the total MSD. With the phonon branches being orthogonal, this is given by
\begin{equation}
\langle u^2 \rangle = \sum_{i=1}^3 \langle u^2_i \rangle.
\end{equation}

Figure \ref{fig:4tm}(a) compares the MSD determined experimentally with the MSD obtained with the optimized solution of the NLM, where $C_e(T_e)$, $C_{p,i}(T_{p,i})$ and{, in particular,} $G_{ep, i}(T_e)$ are taken from the \textit{ab initio} calculation, and $G_{pp}$, $u_E$ and $t_0$ are optimized iteratively. The only free parameter affecting the dynamics of the MSD is $G_{pp}$. We find excellent agreement between data and NLM using a phonon-phonon coupling of approximately $3.5\cdot 10^{17}$\,WK$^{-1}$m$^{-3}$ for all performed measurements. Even though literature values for phonon-phonon coupling are elusive, it has been concluded from neutron-diffraction measurements that electron-phonon and phonon-phonon coupling are of similar magnitude~\cite{Tang2010}, which is consistent with our observation.

Figure~\ref{fig:4tm}(b) shows $T_e$ and $T_{p,i}$ of the optimized solution of the NLM. The different evolution of the temperatures of the three branches is pronounced and shows the transient non-thermal state of the phonons. As expected from the previous discussion, $T_{p,LA}$ rises faster than the temperature of the transverse branches. $T_{p,LA}$ overshoots before the \textit{LA} phonons eventually equilibrate with the two other branches. The timescale of lattice thermalization is sensitive to $G_{pp}$ and is found to be approximately 2\,ps.

\section{Discussion and Conclusion}
\label{sec:conc}

Whereas the TTM and the NLM can both be applied to describe the temporal evolution of the measured MSD with similar quality, the two models differ significantly in the electron-phonon coupling constants. $G_{ep}$ extracted from the data employing the TTM is a factor of two smaller than the value obtained from \textit{ab initio} calculations, i.e., the integrated electron-phonon coupling of all phonon branches. While assuming thermal phonon distributions, which is a basic approximation of the TTM, is inappropriate in non-equilibrium states, the NLM allows for an approximate description of non-thermal phonon distributions.  Our results imply that the determination of the electron-phonon coupling of metals with time-resolved measurements by means of a TTM analysis may lead to systematic errors. In contrast, the excellent agreement between data and NLM employing electron-phonon coupling constants from theory suggests that the NLM represents a minimal model capturing the physics of the electron-phonon coupling in aluminium with sufficient accuracy.

%%%%%%%%%%%%%%%% energy content %%%%%%%%%%%%%%%%%%%%%%%%%%%%%%%%%%%%%%%%%%%%%%%%%%%%%%%%%%%
\begin{figure*}[*htb]
\begin{center}
\includegraphics[width=1.0\textwidth]{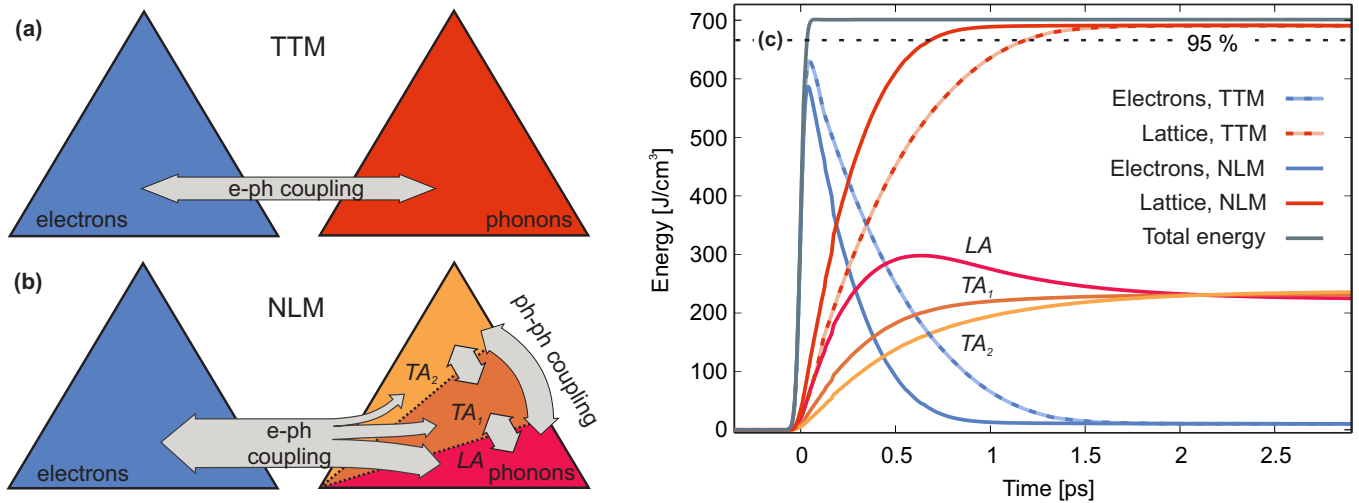}
\caption{(a)\,\&\,(b) Schematic representation of the TTM and the NLM. In the latter, the phononic system is subdivided into its branches and the couplings are considered separately. The widths of the arrows are proportional to the coupling strengths retrieved from theory and experiment. (c) Temporal evolution of the energy content in the electronic (dash-dotted lines) and lattice (solid lines) subsystems for an excitation density of 680\,J/cm$^3$ according to the NLM and the TTM. The energy content of the individual phonon branches is shown as dotted lines. The black dashed line indicates 95\% of the energy which is eventually transferred from electrons to the lattice.}
\label{fig:energy}
\end{center}
\end{figure*}
%%%%%%%%%%%%%%%%%%%%%%%%%%%%%%%%%%%%%%%%%%%%%%%%%%%%%%%%%%%

The important practical difference between the models is the predicted dynamics of the energy flow between electrons and phonons. Figure \ref{fig:energy}(a)\,\&\,(b) visually summarizes the two-temperature and non-thermal lattice model and shows the channels of energy flow. The widths of the arrows indicate the effective coupling strengths. Figure \ref{fig:energy}(c) shows the evolution of the energy content in electrons and phonons added by the laser pulse after excitation of the electrons at $t=0$. The energy content is calculated from the electronic and phononic temperatures given by the two models and the respective heat capacities. In the NLM, the energies of the three phonon branches are summed to obtain the added energy in all lattice degrees of freedom.

It can be seen that the total energy transfer from electrons to the lattice proceeds significantly faster in the NLM. The dotted line depicts 95\% of the energy, which is eventually transferred to the lattice. Between the two models, the time at which this level is reached differs by 40-60\% depending on the excitation density. While the TTM is perfectly suitable for describing the time-resolved diffraction data, as shown in Fig. \ref{fig:TTM}(a), it returns an 'effective' electron-phonon coupling constant which underestimates the rate of energy flow between electrons and phonons approximately by a factor two. This has implications wherever the electron-phonon coupling is used to predict the energy flow between the subsystems, for instance in non-equilibrium MD simulations, the modeling of laser-induced ablation, shock wave and plasma generation, as well as in the description of ultrafast magnetization dynamics, where the energy flow between electrons, lattice and spins is decisive \cite{Beaurepaire1996}. In addition, the few-100\,fs timescale of energy transfer in aluminium implies that photo-induced structural dynamics, e.g.~laser-induced phase transitions, may occur on the sub-ps timescale purely due to incoherent energy transfer. We note that non-thermal electron-distributions might need to be considered additionally for materials exhibiting weak electron-electron scattering~\cite{Mueller2013}.

The refinement of the description of electron-phonon coupling including phonon-specificity is not only relevant for the evolution of impulsively generated highly non-equilibrium states~\cite{Siwick2003, Leguay2013}. Equally important, in quasi-stationary non-equilibrium situation like the flow of electrical current, energy dissipation from the electrons to the lattice is governed by the same coupling phenomena. In other words, the phonon branch-dependence of the electron-phonon coupling may also be applied to the description of electron transport with a Boltzmann equation~\cite{Allen1978}.

The experimental basis of this work is the time-resolved measurement of the atomic mean-square displacement caused by all phonons of aluminium. The energy transfer from electrons to phonons is then inferred on the basis of two models, as it is not an observable itself. However, a more detailed view on transient phonon distributions could be obtained by time- and momentum-resolved measurements of inelastically scattered electrons~\cite{Chase2016} or x-rays~\cite{Trigo2010} from single-crystalline thin films of aluminium. Such studies can provide direct experimental evidence of transient non-thermal phonon distributions, as very recently demonstrated for the nobel metal gold \cite{Chase2016}. 

Heterogeneous coupling of electrons to different types of phonons is well established for materials exhibiting strongly coupled optical phonons~\cite{Schafer2011, Chatelain2014, KST2003, Zijlstra2006, Bothschafter2013, Giret2011, Arnaud2013}. Our work indicates that phonon-specific coupling has to be considered even in simple crystals lacking optical phonons like the free-electron metal aluminium. 

The concept of subdividing the lattice degrees of freedom into individual phonon branches for describing the energy flow in a material, as introduced here for the case of aluminium, may be applied to a range of materials. In the case of metals, a pronounced spectral dependence of $\alpha^2(\omega)$ suggests that this approach is required, and we expect partitioning into phonon branches to be a generally suitable approach. More complex materials may require partitioning into a larger number of phononic subsystems. \textit{Ab initio} calculations can guide in identifying the appropriate subsystems and provide the partial coupling constants, as demonstrated here.

\section{acknowledgments}
This work was funded by the Max Planck Society. L.W.~acknowledges support by the Leibniz graduate school 'Dynamics in New Light'. R.B.~thanks the Alexander von Humboldt Foundation for financial support. J.V.~thanks Prof.~S.H.~Glenzer and SIMES for generous hospitality. The authors thank A.~Paarmann and T.~Kampfrath for carefully reading the manuscript.

\appendix*
\section{Computational details}
\label{sec:app}
All simulations were performed with the \textit{ab initio} simulation package {\em abinit} \cite{abinit,Gonze:2009}.
The electron-phonon coupling calculations capabilities are described in Refs. \cite{Gonze:1997a,Gonze:1997b} and are implemented similarly to the one in Ref. \cite{Savrasov:1996}.
We used a Troullier-Martins type norm conserving pseudopotential from the fhi database with the outermost three electrons taken into account explicitly and the remaining 10 electrons frozen in the core \cite{Fuchs:1999}. The planewave cutoff was $E_{cut}=13$\,Ha and the maximum angular quantum number is $l=2$.
We checked the results for the electronic DOS and the Eliashberg function {with a calculation featuring} a harder pseudopotential requiring a plane wave cutoff of $40$\,Ha {and confirmed our results}. The exchange correlation potential was taken in the generalized gradient approximation (PBE-GGA) \cite{perdew:1996}. Additionally, we performed a calculation in local density approximation (LDA), which gave a slightly different shape of the Eliashberg function in phonon frequency space but left the first moment of the Eliashberg function unchanged. We took the experimental lattice constant for fcc aluminium $a=7.652$\,a$_B$. The k-point grid was unshifted featuring $32\times 32\times 32$ points. The q-point grid was commensurable with the k-point grid at $8\times 8 \times 8$ points. The occupation of the eigenvalues was determined by a Fermi distribution with $0.03$\,eV smearing.
%%%%%%%%%%%%%%%% appendix: eDOS %%%%%%%%%%%%%%%%%%%%%%%%%%%%%%%%%%%%%%%%%%%%%%%%%%%%%%%%%%%
\begin{figure}[tbh]
\begin{center}
\includegraphics[width=0.95\columnwidth]{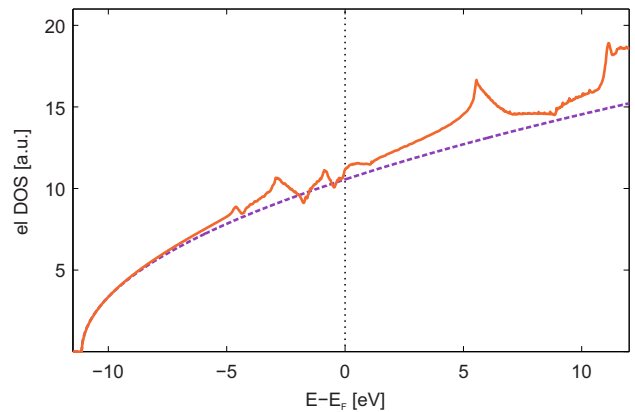}
\caption{DFT calculation of the electronic density of states in aluminium. The dashed line illustrates the $\sqrt{E}$ behaviour of a free e-gas.}
\label{fig:eDOS}
\end{center}
\end{figure}
%%%%%%%%%%%%%%%%%%%%%%%%%%%%%%%%%%%%%%%%%%%%%%%%%%%%%%%%%%%

%%%%%%%%%%%%%%%% appendix: partial a2F and DOS%%%%%%%%%%%%%%%%%%%%%%%%%%%%%%%%%%%%%%%%
\begin{figure}[tbh]
\begin{center}
\includegraphics[width=0.95\columnwidth]{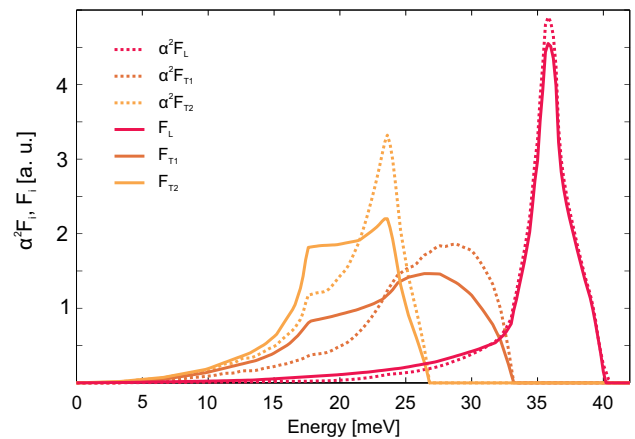}
\caption{Comparison of the spectral shapes of the partial Eliashberg functions $\alpha^2 F_i$ and partial phonon density of states $F_i$. Each function is normalized to its integral.}
\label{fig:partial}
\end{center}
\end{figure}
%%%%%%%%%%%%%%%%%%%%%%%%%%%%%%%%%%%%%%%%%%%%%%%%%%%%%%%%%%%

Such a calculation was performed for the unit cell of fcc aluminium in order to obtain the values for the Eliashberg function. Similarly, the electronic DOS was obtained using the tetrahedron method featuring a k-point grid up to $128\times 128 \times 128$ for the ideal lattice at $T_l=0$\,K. Electronic temperature effects on the DOS and the Eliashberg function up to $T_e=20000$\,K have been checked and found to be small. In order to analyze lattice temperature effects on the electron-phonon energy transfer, we prepared density functional molecular dynamics (DFT-MD) runs with $N_i=32$ aluminium atoms in a super cell at normal density of $\rho=2.7$\,g/cc. The MD part used a Verlet algorithm and Nos\'e thermostat with a time step of $\Delta t =0.2$\,fs. We allowed the ions in the supercell to equilibrate over several thousand time steps at $T_l=300$\,K, $T_l=500$\,K, $T_l=700$\,K, $T_l=900$\,K, and $T_l=1100$\,K. From the equilibrated part of the runs representing fcc lattices with irregular harmonic and anharmonic lattice distortions, we randomly took 10 snapshots each and subjected these snapshots to the same electron-phonon calculations as the initial unit cell. As usual, three perturbation directions per atom have to be considered in order to {build} the electron-phonon matrix elements, requiring 96 separate calculations per temperature and snapshot. In addition, due to the size of supercell and constraints in the available computer memory, we chose to reduce the k-point grid for these cases to $4\times 4 \times 4$ and the q-point grid to $2\times 2\times 2$. We also ran simulations with $8\times 8 \times 8$ k-points to check on the error introduced this way {and found only very minor deviations in the Eliashberg function}. As the electronic DOS in aluminium is already close to the one for an ideal free electron gas, we observed only small changes in the DOS with an increase in lattice temperature. Using the electronic DOS we can also obtain the electronic heat capacity needed in the TTM \cite{Lin2008}. The lattice heat capacity is obtained using the same phonons as used for the calculation of the electron-phonon coupling.

The partial Eliasberg functions for the NLM can be obtained from the phonon line-widths and the partial phonon DOS as computed with abinit. A practical and simple approximation is given by dividing the total Eliashberg function by the total Phonon DOS and using the quotient together with the partial phonon DOS to obtain partial Eliashberg functions. In \fig{fig:partial} the partial phonon DOS $F_i$ and the so-obtained partial Eliashberg functions $\alpha^2F(\omega)$, normalized by their respective integral, are plotted.  

The difference in $G_{ep}$ between our result and the one from Lin {\em et al.}~\cite{Lin2008}, see \fig{fig:TTM}, is caused by two different reasons. Firstly, there is the difference in the electronic DOS. The pronounced oscillations in the DOS in Ref.~\cite{Lin2008}, in particular near the Fermi level, cause the strong rise of $G_{ep}(T_e)$ for $T_e$ increasing from 0 to 5000~K. As such features are absent in the DOS of this work, which is in very good agreement with other \textit{ab initio} calculations \cite{Smrcka1970, Levinson1983}, the $T_e$-dependence of $G_{ep}$ is weak. We note that the DOS of aluminium of Ref.~\cite{Lin2008} is reproduced by the tetrahedron method with $11\times 11\times 11$ k-points. Secondly, and more importantly, there is a difference in the value for the moment of the Eliashberg function $\lambda\langle\omega^2 \rangle$. Our result $\lambda\langle \omega^2\rangle=4.81\cdot 10^{-7}$\,Ha$^2$, $\lambda=0.47$ is obtained entirely on the basis of our DFT calculations and agrees well with other DFT calculations ($\lambda=0.43$ \cite{Nechaev2008}), ($\lambda\langle \omega^2\rangle=4.79\cdot 10^{-7}$\,Ha$^2$, $\lambda=0.44$) and experimental results ($\lambda\langle \omega^2\rangle=5.1\cdot 10^{-7}$\,Ha$^2$, $\lambda=0.42$) given by Savrasov {\em et al.}~\cite{Savrasov:1996}. In addition, our prediction for the superconductivity temperature $T_c$ as obtained from the values for the electron-phonon coupling via the McMillan formula \cite{McMillan:1968} is $T_c=1.35$\,K compared to the experimental value of $T_c=1.2$\,K. The value of $\lambda\langle\omega^2\rangle=2.51\cdot 10^{-7}$\,Ha$^2$ used by Lin {\em et al.} is not obtained from DFT but is based on experimental results {employing a TTM in the data analysis}~\cite{Hostetler1999} and uses the approximation $\langle\omega^2\rangle=\theta_D^2/2$, with $\theta_D$ being the experimental Debye temperature.

\bibliography{Al_literatur}

%merlin.mbs apsrev4-1.bst 2010-07-25 4.21a (PWD, AO, DPC) hacked
%Control: key (0)
%Control: author (0) dotless jnrlst
%Control: editor formatted (1) identically to author
%Control: production of article title (0) allowed
%Control: page (1) range
%Control: year (0) verbatim
%Control: production of eprint (0) enabled
\begin{thebibliography}{53}%
\makeatletter
\providecommand \@ifxundefined [1]{%
 \@ifx{#1\undefined}
}%
\providecommand \@ifnum [1]{%
 \ifnum #1\expandafter \@firstoftwo
 \else \expandafter \@secondoftwo
 \fi
}%
\providecommand \@ifx [1]{%
 \ifx #1\expandafter \@firstoftwo
 \else \expandafter \@secondoftwo
 \fi
}%
\providecommand \natexlab [1]{#1}%
\providecommand \enquote  [1]{``#1''}%
\providecommand \bibnamefont  [1]{#1}%
\providecommand \bibfnamefont [1]{#1}%
\providecommand \citenamefont [1]{#1}%
\providecommand \href@noop [0]{\@secondoftwo}%
\providecommand \href [0]{\begingroup \@sanitize@url \@href}%
\providecommand \@href[1]{\@@startlink{#1}\@@href}%
\providecommand \@@href[1]{\endgroup#1\@@endlink}%
\providecommand \@sanitize@url [0]{\catcode `\\12\catcode `\$12\catcode
  `\&12\catcode `\#12\catcode `\^12\catcode `\_12\catcode `\%12\relax}%
\providecommand \@@startlink[1]{}%
\providecommand \@@endlink[0]{}%
\providecommand \url  [0]{\begingroup\@sanitize@url \@url }%
\providecommand \@url [1]{\endgroup\@href {#1}{\urlprefix }}%
\providecommand \urlprefix  [0]{URL }%
\providecommand \Eprint [0]{\href }%
\providecommand \doibase [0]{http://dx.doi.org/}%
\providecommand \selectlanguage [0]{\@gobble}%
\providecommand \bibinfo  [0]{\@secondoftwo}%
\providecommand \bibfield  [0]{\@secondoftwo}%
\providecommand \translation [1]{[#1]}%
\providecommand \BibitemOpen [0]{}%
\providecommand \bibitemStop [0]{}%
\providecommand \bibitemNoStop [0]{.\EOS\space}%
\providecommand \EOS [0]{\spacefactor3000\relax}%
\providecommand \BibitemShut  [1]{\csname bibitem#1\endcsname}%
\let\auto@bib@innerbib\@empty
%</preamble>
\bibitem [{\citenamefont {Grimvall}\ and\ \citenamefont
  {Wohlfahrt}(1981)}]{book:Grimvall}%
  \BibitemOpen
  \bibinfo {editor} {\bibfnamefont {G.}~\bibnamefont {Grimvall}}\ and\ \bibinfo
  {editor} {\bibfnamefont {E.~P.}\ \bibnamefont {Wohlfahrt}},\ eds.,\
  \href@noop {} {\emph {\bibinfo {title} {The Electron-Phonon Interaction in
  Metals}}},\ \bibinfo {series} {Selected Topics in Solid State Physics},
  Vol.~\bibinfo {volume} {16}\ (\bibinfo  {publisher} {North-Holland
  publishing},\ \bibinfo {year} {1981})\BibitemShut {NoStop}%
\bibitem [{\citenamefont {Patterson}\ and\ \citenamefont
  {Bailey}(2010)}]{book:Patterson}%
  \BibitemOpen
  \bibinfo {editor} {\bibfnamefont {J.}~\bibnamefont {Patterson}}\ and\
  \bibinfo {editor} {\bibfnamefont {B.}~\bibnamefont {Bailey}},\ eds.,\
  \href@noop {} {\emph {\bibinfo {title} {Solid-State Physics}}}\ (\bibinfo
  {publisher} {Springer Verlag Berlin Heidelberg},\ \bibinfo {year}
  {2010})\BibitemShut {NoStop}%
\bibitem [{\citenamefont {Anisimov}\ \emph {et~al.}(1967)\citenamefont
  {Anisimov}, \citenamefont {Bonch-Bruevich}, \citenamefont {El'yashevich},
  \citenamefont {Imas}, \citenamefont {Pavlenko},\ and\ \citenamefont
  {Romanov}}]{Anisimov1967}%
  \BibitemOpen
  \bibfield  {author} {\bibinfo {author} {\bibfnamefont {S.~I.}\ \bibnamefont
  {Anisimov}}, \bibinfo {author} {\bibfnamefont {A.~M.}\ \bibnamefont
  {Bonch-Bruevich}}, \bibinfo {author} {\bibfnamefont {M.~A.}\ \bibnamefont
  {El'yashevich}}, \bibinfo {author} {\bibfnamefont {Ya.~A.}\ \bibnamefont
  {Imas}}, \bibinfo {author} {\bibfnamefont {N.~A.}\ \bibnamefont {Pavlenko}},
  \ and\ \bibinfo {author} {\bibfnamefont {G.~S.}\ \bibnamefont {Romanov}},\
  }\bibfield  {title} {\enquote {\bibinfo {title} {Effect of powerful light
  fluxes on metals},}\ }\href@noop {} {\bibfield  {journal} {\bibinfo
  {journal} {Sov. Phys. Tech. Phys.}\ }\textbf {\bibinfo {volume} {11}},\
  \bibinfo {pages} {945} (\bibinfo {year} {1967})}\BibitemShut {NoStop}%
\bibitem [{\citenamefont {Anisimov}\ \emph {et~al.}(1974)\citenamefont
  {Anisimov}, \citenamefont {Kapeliovich},\ and\ \citenamefont
  {Perel'man}}]{Anisimov1974}%
  \BibitemOpen
  \bibfield  {author} {\bibinfo {author} {\bibfnamefont {S.~I.}\ \bibnamefont
  {Anisimov}}, \bibinfo {author} {\bibfnamefont {B.~L.}\ \bibnamefont
  {Kapeliovich}}, \ and\ \bibinfo {author} {\bibfnamefont {T.~L.}\ \bibnamefont
  {Perel'man}},\ }\bibfield  {title} {\enquote {\bibinfo {title} {{Electron
  emission from metal surfaces exposed to ultrashort laser pulses}},}\
  }\href@noop {} {\bibfield  {journal} {\bibinfo  {journal} {Zh. Eksp. Teor.
  Fiz.}\ }\textbf {\bibinfo {volume} {66}},\ \bibinfo {pages} {776--781}
  (\bibinfo {year} {1974})}\BibitemShut {NoStop}%
\bibitem [{\citenamefont {Allen}(1987)}]{Allen1987}%
  \BibitemOpen
  \bibfield  {author} {\bibinfo {author} {\bibfnamefont {P.~B.}\ \bibnamefont
  {Allen}},\ }\bibfield  {title} {\enquote {\bibinfo {title} {{Theory of
  thermal relaxation of electrons in metals}},}\ }\href
  {http://journals.aps.org/prl/abstract/10.1103/PhysRevLett.59.1460} {\bibfield
   {journal} {\bibinfo  {journal} {Physical Review Letters}\ }\textbf {\bibinfo
  {volume} {59}},\ \bibinfo {pages} {1460--1463} (\bibinfo {year}
  {1987})}\BibitemShut {NoStop}%
\bibitem [{\citenamefont {Lin}\ \emph {et~al.}(2008)\citenamefont {Lin},
  \citenamefont {Zhigilei},\ and\ \citenamefont {Celli}}]{Lin2008}%
  \BibitemOpen
  \bibfield  {author} {\bibinfo {author} {\bibfnamefont {Z.}~\bibnamefont
  {Lin}}, \bibinfo {author} {\bibfnamefont {L.~V.}\ \bibnamefont {Zhigilei}}, \
  and\ \bibinfo {author} {\bibfnamefont {V.}~\bibnamefont {Celli}},\ }\bibfield
   {title} {\enquote {\bibinfo {title} {{Electron-phonon coupling and electron
  heat capacity of metals under conditions of strong electron-phonon
  nonequilibrium}},}\ }\href {\doibase 10.1103/PhysRevB.77.075133} {\bibfield
  {journal} {\bibinfo  {journal} {Physical Review B}\ }\textbf {\bibinfo
  {volume} {77}},\ \bibinfo {pages} {075133} (\bibinfo {year}
  {2008})}\BibitemShut {NoStop}%
\bibitem [{\citenamefont {Rethfeld}\ \emph {et~al.}(2002)\citenamefont
  {Rethfeld}, \citenamefont {Kaiser}, \citenamefont {Vicanek},\ and\
  \citenamefont {Simon}}]{Rethfeld2002}%
  \BibitemOpen
  \bibfield  {author} {\bibinfo {author} {\bibfnamefont {B.}~\bibnamefont
  {Rethfeld}}, \bibinfo {author} {\bibfnamefont {A.}~\bibnamefont {Kaiser}},
  \bibinfo {author} {\bibfnamefont {M.}~\bibnamefont {Vicanek}}, \ and\
  \bibinfo {author} {\bibfnamefont {G.}~\bibnamefont {Simon}},\ }\bibfield
  {title} {\enquote {\bibinfo {title} {{Ultrafast dynamics of nonequilibrium
  electrons in metals under femtosecond laser irradiation}},}\ }\href {\doibase
  10.1103/PhysRevB.65.214303} {\bibfield  {journal} {\bibinfo  {journal}
  {Physical Review B}\ }\textbf {\bibinfo {volume} {65}},\ \bibinfo {pages}
  {214303} (\bibinfo {year} {2002})}\BibitemShut {NoStop}%
\bibitem [{\citenamefont {Stedman}\ \emph {et~al.}(1967)\citenamefont
  {Stedman}, \citenamefont {Almqvist},\ and\ \citenamefont
  {Nilsson}}]{Stedman1967}%
  \BibitemOpen
  \bibfield  {author} {\bibinfo {author} {\bibfnamefont {R.}~\bibnamefont
  {Stedman}}, \bibinfo {author} {\bibfnamefont {L.}~\bibnamefont {Almqvist}}, \
  and\ \bibinfo {author} {\bibfnamefont {G.}~\bibnamefont {Nilsson}},\
  }\bibfield  {title} {\enquote {\bibinfo {title} {{Phonon-frequency
  distributions and heat capacities of aluminum and lead}},}\ }\href
  {http://journals.aps.org/pr/abstract/10.1103/PhysRev.162.549} {\bibfield
  {journal} {\bibinfo  {journal} {Physical Review}\ }\textbf {\bibinfo {volume}
  {162}},\ \bibinfo {pages} {549--557} (\bibinfo {year} {1967})}\BibitemShut
  {NoStop}%
\bibitem [{\citenamefont {McMillan}\ and\ \citenamefont
  {Rowell}(1965)}]{McMillan1965}%
  \BibitemOpen
  \bibfield  {author} {\bibinfo {author} {\bibfnamefont {W.~L.}\ \bibnamefont
  {McMillan}}\ and\ \bibinfo {author} {\bibfnamefont {J.~M.}\ \bibnamefont
  {Rowell}},\ }\bibfield  {title} {\enquote {\bibinfo {title} {{Lead phonon
  spectrum calculated from superconducting density of states}},}\ }\href
  {\doibase 10.1103/PhysRevLett.14.108} {\bibfield  {journal} {\bibinfo
  {journal} {Physical Review Letters}\ }\textbf {\bibinfo {volume} {14}},\
  \bibinfo {pages} {108--112} (\bibinfo {year} {1965})}\BibitemShut {NoStop}%
\bibitem [{\citenamefont {Plummer}\ \emph {et~al.}(2003)\citenamefont
  {Plummer}, \citenamefont {Shi}, \citenamefont {Tang}, \citenamefont
  {Rotenberg},\ and\ \citenamefont {Kevan}}]{Plummer2003}%
  \BibitemOpen
  \bibfield  {author} {\bibinfo {author} {\bibfnamefont {E.~W.}\ \bibnamefont
  {Plummer}}, \bibinfo {author} {\bibfnamefont {J.}~\bibnamefont {Shi}},
  \bibinfo {author} {\bibfnamefont {S.-J.}\ \bibnamefont {Tang}}, \bibinfo
  {author} {\bibfnamefont {E.}~\bibnamefont {Rotenberg}}, \ and\ \bibinfo
  {author} {\bibfnamefont {S.~D.}\ \bibnamefont {Kevan}},\ }\bibfield  {title}
  {\enquote {\bibinfo {title} {{Enhanced electron-phonon coupling at metal
  surfaces}},}\ }\href {\doibase 10.1016/j.progsurf.2003.08.033} {\bibfield
  {journal} {\bibinfo  {journal} {Progress in Surface Science}\ }\textbf
  {\bibinfo {volume} {74}},\ \bibinfo {pages} {251--268} (\bibinfo {year}
  {2003})}\BibitemShut {NoStop}%
\bibitem [{\citenamefont {Elsayed-Ali}\ \emph {et~al.}(1987)\citenamefont
  {Elsayed-Ali}, \citenamefont {Norris}, \citenamefont {Pessot},\ and\
  \citenamefont {Mourou}}]{Elsayed-Ali1987}%
  \BibitemOpen
  \bibfield  {author} {\bibinfo {author} {\bibfnamefont {H.~E.}\ \bibnamefont
  {Elsayed-Ali}}, \bibinfo {author} {\bibfnamefont {T.~B.}\ \bibnamefont
  {Norris}}, \bibinfo {author} {\bibfnamefont {M.~A.}\ \bibnamefont {Pessot}},
  \ and\ \bibinfo {author} {\bibfnamefont {G.~A.}\ \bibnamefont {Mourou}},\
  }\bibfield  {title} {\enquote {\bibinfo {title} {{Time-resolved observation
  of electron-phonon relaxation in copper}},}\ }\href
  {http://journals.aps.org/prl/abstract/10.1103/PhysRevLett.58.1212} {\bibfield
   {journal} {\bibinfo  {journal} {Physical Review Letters}\ }\textbf {\bibinfo
  {volume} {58}},\ \bibinfo {pages} {1212--1215} (\bibinfo {year}
  {1987})}\BibitemShut {NoStop}%
\bibitem [{\citenamefont {Schoenlein}\ \emph {et~al.}(1987)\citenamefont
  {Schoenlein}, \citenamefont {Lin}, \citenamefont {Fujimoto},\ and\
  \citenamefont {Eesley}}]{Schoenlein1987}%
  \BibitemOpen
  \bibfield  {author} {\bibinfo {author} {\bibfnamefont {R.~W.}\ \bibnamefont
  {Schoenlein}}, \bibinfo {author} {\bibfnamefont {W.~Z.}\ \bibnamefont {Lin}},
  \bibinfo {author} {\bibfnamefont {J.~G.}\ \bibnamefont {Fujimoto}}, \ and\
  \bibinfo {author} {\bibfnamefont {G.~L.}\ \bibnamefont {Eesley}},\ }\bibfield
   {title} {\enquote {\bibinfo {title} {{Femtosecond studies of nonequilibrium
  electronic processes in metals}},}\ }\href
  {http://journals.aps.org/prl/abstract/10.1103/PhysRevLett.58.1680} {\bibfield
   {journal} {\bibinfo  {journal} {Physical Review Letters}\ }\textbf {\bibinfo
  {volume} {58}},\ \bibinfo {pages} {1680--1683} (\bibinfo {year}
  {1987})}\BibitemShut {NoStop}%
\bibitem [{\citenamefont {Brorson}\ \emph {et~al.}(1990)\citenamefont
  {Brorson}, \citenamefont {Kazeroonian}, \citenamefont {Moodera},
  \citenamefont {Face}, \citenamefont {Cheng}, \citenamefont {Ippen},
  \citenamefont {Dresselhaus},\ and\ \citenamefont
  {Dresselhaus}}]{Brorson1990}%
  \BibitemOpen
  \bibfield  {author} {\bibinfo {author} {\bibfnamefont {S.~D.}\ \bibnamefont
  {Brorson}}, \bibinfo {author} {\bibfnamefont {A.}~\bibnamefont
  {Kazeroonian}}, \bibinfo {author} {\bibfnamefont {J.~S.}\ \bibnamefont
  {Moodera}}, \bibinfo {author} {\bibfnamefont {D.~W.}\ \bibnamefont {Face}},
  \bibinfo {author} {\bibfnamefont {T.~K.}\ \bibnamefont {Cheng}}, \bibinfo
  {author} {\bibfnamefont {E.~P.}\ \bibnamefont {Ippen}}, \bibinfo {author}
  {\bibfnamefont {M.~S.}\ \bibnamefont {Dresselhaus}}, \ and\ \bibinfo {author}
  {\bibfnamefont {G.}~\bibnamefont {Dresselhaus}},\ }\bibfield  {title}
  {\enquote {\bibinfo {title} {{Femtosecond room-temperature measurement of the
  electron-phonon coupling constant $\gamma$ in metallic superconductors}},}\
  }\href {\doibase 10.1103/PhysRevLett.64.2172} {\bibfield  {journal} {\bibinfo
   {journal} {Physical Review Letters}\ }\textbf {\bibinfo {volume} {64}},\
  \bibinfo {pages} {2172--2175} (\bibinfo {year} {1990})}\BibitemShut {NoStop}%
\bibitem [{\citenamefont {Sentef}\ \emph {et~al.}(2013)\citenamefont {Sentef},
  \citenamefont {Kemper}, \citenamefont {Moritz}, \citenamefont {Freericks},
  \citenamefont {Shen},\ and\ \citenamefont {Devereaux}}]{Sentef2013}%
  \BibitemOpen
  \bibfield  {author} {\bibinfo {author} {\bibfnamefont {M.}~\bibnamefont
  {Sentef}}, \bibinfo {author} {\bibfnamefont {A.~F.}\ \bibnamefont {Kemper}},
  \bibinfo {author} {\bibfnamefont {B.}~\bibnamefont {Moritz}}, \bibinfo
  {author} {\bibfnamefont {J.~K.}\ \bibnamefont {Freericks}}, \bibinfo {author}
  {\bibfnamefont {Z.-X.}\ \bibnamefont {Shen}}, \ and\ \bibinfo {author}
  {\bibfnamefont {T.~P.}\ \bibnamefont {Devereaux}},\ }\bibfield  {title}
  {\enquote {\bibinfo {title} {{Examining Electron-Boson Coupling Using
  Time-Resolved Spectroscopy}},}\ }\href {\doibase 10.1103/PhysRevX.3.041033}
  {\bibfield  {journal} {\bibinfo  {journal} {Physical Review X}\ }\textbf
  {\bibinfo {volume} {3}},\ \bibinfo {pages} {041033} (\bibinfo {year}
  {2013})}\BibitemShut {NoStop}%
\bibitem [{\citenamefont {Bauer}\ \emph {et~al.}(1998)\citenamefont {Bauer},
  \citenamefont {Pawlik},\ and\ \citenamefont {Aeschlimann}}]{Bauer1998}%
  \BibitemOpen
  \bibfield  {author} {\bibinfo {author} {\bibfnamefont {M.}~\bibnamefont
  {Bauer}}, \bibinfo {author} {\bibfnamefont {S.}~\bibnamefont {Pawlik}}, \
  and\ \bibinfo {author} {\bibfnamefont {M.}~\bibnamefont {Aeschlimann}},\
  }\bibfield  {title} {\enquote {\bibinfo {title} {{Electron dynamics of
  aluminum investigated by means of time-resolved photoemission}},}\ }\href
  {\doibase 10.1117/12.307123} {\bibfield  {journal} {\bibinfo  {journal}
  {SPIE}\ }\textbf {\bibinfo {volume} {3272}},\ \bibinfo {pages} {201--210}
  (\bibinfo {year} {1998})}\BibitemShut {NoStop}%
\bibitem [{\citenamefont {Hopkins}\ and\ \citenamefont
  {Norris}(2009)}]{Hopkins2009}%
  \BibitemOpen
  \bibfield  {author} {\bibinfo {author} {\bibfnamefont {P.~E.}\ \bibnamefont
  {Hopkins}}\ and\ \bibinfo {author} {\bibfnamefont {P.~M.}\ \bibnamefont
  {Norris}},\ }\bibfield  {title} {\enquote {\bibinfo {title} {{Contribution of
  Ballistic Electron Transport to Energy Transfer During Electron-Phonon
  Nonequilibrium in Thin Metal Films}},}\ }\href {\doibase 10.1115/1.3072929}
  {\bibfield  {journal} {\bibinfo  {journal} {Journal of Heat Transfer}\
  }\textbf {\bibinfo {volume} {131}},\ \bibinfo {pages} {043208} (\bibinfo
  {year} {2009})}\BibitemShut {NoStop}%
\bibitem [{\citenamefont {Williamson}\ \emph {et~al.}(1984)\citenamefont
  {Williamson}, \citenamefont {Mourou},\ and\ \citenamefont
  {Li}}]{Williamson1984}%
  \BibitemOpen
  \bibfield  {author} {\bibinfo {author} {\bibfnamefont {S.}~\bibnamefont
  {Williamson}}, \bibinfo {author} {\bibfnamefont {G.}~\bibnamefont {Mourou}},
  \ and\ \bibinfo {author} {\bibfnamefont {J.~C.~M.}\ \bibnamefont {Li}},\
  }\bibfield  {title} {\enquote {\bibinfo {title} {{Time-resolved,
  laser-induced phase transformation in aluminum}},}\ }\href
  {https://journals.aps.org/prl/pdf/10.1103/PhysRevLett.52.2364} {\bibfield
  {journal} {\bibinfo  {journal} {Physical Review Letters}\ }\textbf {\bibinfo
  {volume} {52}},\ \bibinfo {pages} {2364} (\bibinfo {year}
  {1984})}\BibitemShut {NoStop}%
\bibitem [{\citenamefont {Siwick}\ \emph {et~al.}(2003)\citenamefont {Siwick},
  \citenamefont {Dwyer}, \citenamefont {Jordan},\ and\ \citenamefont
  {Miller}}]{Siwick2003}%
  \BibitemOpen
  \bibfield  {author} {\bibinfo {author} {\bibfnamefont {B.~J.}\ \bibnamefont
  {Siwick}}, \bibinfo {author} {\bibfnamefont {J.~R.}\ \bibnamefont {Dwyer}},
  \bibinfo {author} {\bibfnamefont {R.~E.}\ \bibnamefont {Jordan}}, \ and\
  \bibinfo {author} {\bibfnamefont {R.~J.~D.}\ \bibnamefont {Miller}},\
  }\bibfield  {title} {\enquote {\bibinfo {title} {{An atomic-level view of
  melting using femtosecond electron diffraction.}}}\ }\href {\doibase
  10.1126/science.1090052} {\bibfield  {journal} {\bibinfo  {journal}
  {Science}\ }\textbf {\bibinfo {volume} {302}},\ \bibinfo {pages} {1382--1385}
  (\bibinfo {year} {2003})}\BibitemShut {NoStop}%
\bibitem [{\citenamefont {Waldecker}\ \emph {et~al.}(2015)\citenamefont
  {Waldecker}, \citenamefont {Bertoni},\ and\ \citenamefont
  {Ernstorfer}}]{Waldecker2015Setup}%
  \BibitemOpen
  \bibfield  {author} {\bibinfo {author} {\bibfnamefont {L.}~\bibnamefont
  {Waldecker}}, \bibinfo {author} {\bibfnamefont {R.}~\bibnamefont {Bertoni}},
  \ and\ \bibinfo {author} {\bibfnamefont {R.}~\bibnamefont {Ernstorfer}},\
  }\bibfield  {title} {\enquote {\bibinfo {title} {{Compact femtosecond
  electron diffractometer with 100 keV electron bunches approaching the
  single-electron pulse duration limit}},}\ }\href {\doibase 10.1063/1.4906786}
  {\bibfield  {journal} {\bibinfo  {journal} {Journal of Applied Physics}\
  }\textbf {\bibinfo {volume} {117}},\ \bibinfo {pages} {044903} (\bibinfo
  {year} {2015})}\BibitemShut {NoStop}%
\bibitem [{\citenamefont {Peng}\ \emph {et~al.}(2004)\citenamefont {Peng},
  \citenamefont {Dudarev},\ and\ \citenamefont {Whelan}}]{book:Peng}%
  \BibitemOpen
  \bibfield  {author} {\bibinfo {author} {\bibfnamefont {L.~M.}\ \bibnamefont
  {Peng}}, \bibinfo {author} {\bibfnamefont {S.~L.}\ \bibnamefont {Dudarev}}, \
  and\ \bibinfo {author} {\bibfnamefont {M.~J.}\ \bibnamefont {Whelan}},\
  }\href@noop {} {\emph {\bibinfo {title} {High-Energy Electron Diffraction and
  Microscopy}}}\ (\bibinfo  {publisher} {Oxford Science Publications},\
  \bibinfo {year} {2004})\BibitemShut {NoStop}%
\bibitem [{\citenamefont {Nie}\ \emph {et~al.}(2006)\citenamefont {Nie},
  \citenamefont {Wang}, \citenamefont {Park}, \citenamefont {Clinite},\ and\
  \citenamefont {Cao}}]{Nie2006}%
  \BibitemOpen
  \bibfield  {author} {\bibinfo {author} {\bibfnamefont {S.}~\bibnamefont
  {Nie}}, \bibinfo {author} {\bibfnamefont {X.}~\bibnamefont {Wang}}, \bibinfo
  {author} {\bibfnamefont {H.}~\bibnamefont {Park}}, \bibinfo {author}
  {\bibfnamefont {R.}~\bibnamefont {Clinite}}, \ and\ \bibinfo {author}
  {\bibfnamefont {J.}~\bibnamefont {Cao}},\ }\bibfield  {title} {\enquote
  {\bibinfo {title} {{Measurement of the Electronic Gr\"{u}neisen Constant
  Using Femtosecond Electron Diffraction}},}\ }\href {\doibase
  10.1103/PhysRevLett.96.025901} {\bibfield  {journal} {\bibinfo  {journal}
  {Physical Review Letters}\ }\textbf {\bibinfo {volume} {96}},\ \bibinfo
  {pages} {025901} (\bibinfo {year} {2006})}\BibitemShut {NoStop}%
\bibitem [{\citenamefont {Zhu}\ \emph {et~al.}(2013)\citenamefont {Zhu},
  \citenamefont {Chen}, \citenamefont {Li}, \citenamefont {Chen}, \citenamefont
  {Cao}, \citenamefont {Sheng},\ and\ \citenamefont {Zhang}}]{Zhu2013}%
  \BibitemOpen
  \bibfield  {author} {\bibinfo {author} {\bibfnamefont {P.}~\bibnamefont
  {Zhu}}, \bibinfo {author} {\bibfnamefont {J.}~\bibnamefont {Chen}}, \bibinfo
  {author} {\bibfnamefont {R.}~\bibnamefont {Li}}, \bibinfo {author}
  {\bibfnamefont {L.}~\bibnamefont {Chen}}, \bibinfo {author} {\bibfnamefont
  {J.}~\bibnamefont {Cao}}, \bibinfo {author} {\bibfnamefont {Z.}~\bibnamefont
  {Sheng}}, \ and\ \bibinfo {author} {\bibfnamefont {J.}~\bibnamefont
  {Zhang}},\ }\bibfield  {title} {\enquote {\bibinfo {title} {{Laser-induced
  short-range disorder in aluminum revealed by ultrafast electron diffuse
  scattering}},}\ }\href {\doibase 10.1063/1.4840355} {\bibfield  {journal}
  {\bibinfo  {journal} {Applied Physics Letters}\ }\textbf {\bibinfo {volume}
  {103}},\ \bibinfo {pages} {231914} (\bibinfo {year} {2013})}\BibitemShut
  {NoStop}%
\bibitem [{\citenamefont {Wang}\ \emph {et~al.}(1994)\citenamefont {Wang},
  \citenamefont {Riffe}, \citenamefont {Lee},\ and\ \citenamefont
  {Downer}}]{Wang:1994}%
  \BibitemOpen
  \bibfield  {author} {\bibinfo {author} {\bibfnamefont {X.~Y.}\ \bibnamefont
  {Wang}}, \bibinfo {author} {\bibfnamefont {D.~M.}\ \bibnamefont {Riffe}},
  \bibinfo {author} {\bibfnamefont {Y.-S.}\ \bibnamefont {Lee}}, \ and\
  \bibinfo {author} {\bibfnamefont {M.~C.}\ \bibnamefont {Downer}},\ }\bibfield
   {title} {\enquote {\bibinfo {title} {Time-resolved electron-temperature
  measurement in a highly excited gold target using femtosecond thermionic
  emission},}\ }\href {\doibase http://dx.doi.org/10.1103/PhysRevB.50.8016}
  {\bibfield  {journal} {\bibinfo  {journal} {Phys. Rev. B}\ }\textbf {\bibinfo
  {volume} {50}},\ \bibinfo {pages} {8016} (\bibinfo {year}
  {1994})}\BibitemShut {NoStop}%
\bibitem [{\citenamefont {Recoules}\ \emph {et~al.}(2006)\citenamefont
  {Recoules}, \citenamefont {Cl\'{e}rouin}, \citenamefont {Z\'{e}rah},
  \citenamefont {Anglade},\ and\ \citenamefont {Mazevet}}]{Recoules2006}%
  \BibitemOpen
  \bibfield  {author} {\bibinfo {author} {\bibfnamefont {V.}~\bibnamefont
  {Recoules}}, \bibinfo {author} {\bibfnamefont {J.}~\bibnamefont
  {Cl\'{e}rouin}}, \bibinfo {author} {\bibfnamefont {G.}~\bibnamefont
  {Z\'{e}rah}}, \bibinfo {author} {\bibfnamefont {P.~M.}\ \bibnamefont
  {Anglade}}, \ and\ \bibinfo {author} {\bibfnamefont {S.}~\bibnamefont
  {Mazevet}},\ }\bibfield  {title} {\enquote {\bibinfo {title} {{Effect of
  Intense Laser Irradiation on the Lattice Stability of Semiconductors and
  Metals}},}\ }\href {\doibase 10.1103/PhysRevLett.96.055503} {\bibfield
  {journal} {\bibinfo  {journal} {Physical Review Letters}\ }\textbf {\bibinfo
  {volume} {96}},\ \bibinfo {pages} {055503} (\bibinfo {year}
  {2006})}\BibitemShut {NoStop}%
\bibitem [{\citenamefont {Mueller}\ and\ \citenamefont
  {Rethfeld}(2013)}]{Mueller2013}%
  \BibitemOpen
  \bibfield  {author} {\bibinfo {author} {\bibfnamefont {B.~Y.}\ \bibnamefont
  {Mueller}}\ and\ \bibinfo {author} {\bibfnamefont {B.}~\bibnamefont
  {Rethfeld}},\ }\bibfield  {title} {\enquote {\bibinfo {title} {{Relaxation
  dynamics in laser-excited metals under nonequilibrium conditions}},}\ }\href
  {\doibase 10.1103/PhysRevB.87.035139} {\bibfield  {journal} {\bibinfo
  {journal} {Phys. Rev. B}\ }\textbf {\bibinfo {volume} {87}},\ \bibinfo
  {pages} {035139} (\bibinfo {year} {2013})}\BibitemShut {NoStop}%
\bibitem [{\citenamefont {Hostetler}\ \emph {et~al.}(1999)\citenamefont
  {Hostetler}, \citenamefont {Smith}, \citenamefont {Czajkowsky},\ and\
  \citenamefont {Norris}}]{Hostetler1999}%
  \BibitemOpen
  \bibfield  {author} {\bibinfo {author} {\bibfnamefont {J.~L.}\ \bibnamefont
  {Hostetler}}, \bibinfo {author} {\bibfnamefont {A.~N.}\ \bibnamefont
  {Smith}}, \bibinfo {author} {\bibfnamefont {D.~M.}\ \bibnamefont
  {Czajkowsky}}, \ and\ \bibinfo {author} {\bibfnamefont {P.~M.}\ \bibnamefont
  {Norris}},\ }\bibfield  {title} {\enquote {\bibinfo {title} {{Measurement of
  the Electron-Phonon Coupling Factor Dependence on Film Thickness and Grain
  Size in Au, Cr, and Al}},}\ }\href {\doibase 10.1364/AO.38.003614} {\bibfield
   {journal} {\bibinfo  {journal} {Applied Optics}\ }\textbf {\bibinfo {volume}
  {38}},\ \bibinfo {pages} {3614--3620} (\bibinfo {year} {1999})}\BibitemShut
  {NoStop}%
\bibitem [{\citenamefont {Sch\"{a}fer}\ \emph {et~al.}(2011)\citenamefont
  {Sch\"{a}fer}, \citenamefont {Liang},\ and\ \citenamefont
  {Zewail}}]{Schafer2011}%
  \BibitemOpen
  \bibfield  {author} {\bibinfo {author} {\bibfnamefont {S.}~\bibnamefont
  {Sch\"{a}fer}}, \bibinfo {author} {\bibfnamefont {W.}~\bibnamefont {Liang}},
  \ and\ \bibinfo {author} {\bibfnamefont {A.~H.}\ \bibnamefont {Zewail}},\
  }\bibfield  {title} {\enquote {\bibinfo {title} {{Primary structural dynamics
  in graphite}},}\ }\href {\doibase 10.1088/1367-2630/13/6/063030} {\bibfield
  {journal} {\bibinfo  {journal} {New Journal of Physics}\ }\textbf {\bibinfo
  {volume} {13}},\ \bibinfo {pages} {063030} (\bibinfo {year}
  {2011})}\BibitemShut {NoStop}%
\bibitem [{\citenamefont {Chatelain}\ \emph {et~al.}(2014)\citenamefont
  {Chatelain}, \citenamefont {Morrison}, \citenamefont {Klarenaar},\ and\
  \citenamefont {Siwick}}]{Chatelain2014}%
  \BibitemOpen
  \bibfield  {author} {\bibinfo {author} {\bibfnamefont {R.~P.}\ \bibnamefont
  {Chatelain}}, \bibinfo {author} {\bibfnamefont {V.~R.}\ \bibnamefont
  {Morrison}}, \bibinfo {author} {\bibfnamefont {B.~L.~M.}\ \bibnamefont
  {Klarenaar}}, \ and\ \bibinfo {author} {\bibfnamefont {B.~J.}\ \bibnamefont
  {Siwick}},\ }\bibfield  {title} {\enquote {\bibinfo {title} {{Coherent and
  Incoherent Electron-Phonon Coupling in Graphite Observed with Radio-Frequency
  Compressed Ultrafast Electron Diffraction}},}\ }\href {\doibase
  10.1103/PhysRevLett.113.235502} {\bibfield  {journal} {\bibinfo  {journal}
  {Physical Review Letters}\ }\textbf {\bibinfo {volume} {113}},\ \bibinfo
  {pages} {235502} (\bibinfo {year} {2014})}\BibitemShut {NoStop}%
\bibitem [{\citenamefont {Sokolowski-Tinten}\ \emph {et~al.}(2003)\citenamefont
  {Sokolowski-Tinten}, \citenamefont {Blome}, \citenamefont {Blums},
  \citenamefont {Cavalleri}, \citenamefont {Dietrich}, \citenamefont
  {Tarasevitch}, \citenamefont {Uschmann}, \citenamefont {F\"{o}rster},
  \citenamefont {Kammler}, \citenamefont {{Horn-von-Hoegen}},\ and\
  \citenamefont {{von der Linde}}}]{KST2003}%
  \BibitemOpen
  \bibfield  {author} {\bibinfo {author} {\bibfnamefont {K.}~\bibnamefont
  {Sokolowski-Tinten}}, \bibinfo {author} {\bibfnamefont {C.}~\bibnamefont
  {Blome}}, \bibinfo {author} {\bibfnamefont {J.}~\bibnamefont {Blums}},
  \bibinfo {author} {\bibfnamefont {A.}~\bibnamefont {Cavalleri}}, \bibinfo
  {author} {\bibfnamefont {C.}~\bibnamefont {Dietrich}}, \bibinfo {author}
  {\bibfnamefont {A.}~\bibnamefont {Tarasevitch}}, \bibinfo {author}
  {\bibfnamefont {I.}~\bibnamefont {Uschmann}}, \bibinfo {author}
  {\bibfnamefont {E.}~\bibnamefont {F\"{o}rster}}, \bibinfo {author}
  {\bibfnamefont {M.}~\bibnamefont {Kammler}}, \bibinfo {author} {\bibfnamefont
  {M.}~\bibnamefont {{Horn-von-Hoegen}}}, \ and\ \bibinfo {author}
  {\bibfnamefont {D.}~\bibnamefont {{von der Linde}}},\ }\bibfield  {title}
  {\enquote {\bibinfo {title} {{Femtosecond X-ray measurement of coherent
  lattice vibrations near the Lindemann stability limit.}}}\ }\href {\doibase
  10.1038/nature01490} {\bibfield  {journal} {\bibinfo  {journal} {Nature}\
  }\textbf {\bibinfo {volume} {422}},\ \bibinfo {pages} {287--289} (\bibinfo
  {year} {2003})}\BibitemShut {NoStop}%
\bibitem [{\citenamefont {Zijlstra}\ \emph {et~al.}(2006)\citenamefont
  {Zijlstra}, \citenamefont {Tatarinova},\ and\ \citenamefont
  {Garcia}}]{Zijlstra2006}%
  \BibitemOpen
  \bibfield  {author} {\bibinfo {author} {\bibfnamefont {E.~S.}\ \bibnamefont
  {Zijlstra}}, \bibinfo {author} {\bibfnamefont {L.~L.}\ \bibnamefont
  {Tatarinova}}, \ and\ \bibinfo {author} {\bibfnamefont {M.~E.}\ \bibnamefont
  {Garcia}},\ }\bibfield  {title} {\enquote {\bibinfo {title} {{Laser-induced
  phonon-phonon interactions in bismuth}},}\ }\href {\doibase
  10.1103/PhysRevB.74.220301} {\bibfield  {journal} {\bibinfo  {journal}
  {Physical Review B}\ }\textbf {\bibinfo {volume} {74}},\ \bibinfo {pages}
  {220301(R)} (\bibinfo {year} {2006})}\BibitemShut {NoStop}%
\bibitem [{\citenamefont {Bothschafter}\ \emph {et~al.}(2013)\citenamefont
  {Bothschafter}, \citenamefont {Paarmann}, \citenamefont {Zijlstra},
  \citenamefont {Karpowicz}, \citenamefont {Garcia}, \citenamefont
  {Kienberger},\ and\ \citenamefont {Ernstorfer}}]{Bothschafter2013}%
  \BibitemOpen
  \bibfield  {author} {\bibinfo {author} {\bibfnamefont {E.~M.}\ \bibnamefont
  {Bothschafter}}, \bibinfo {author} {\bibfnamefont {A.}~\bibnamefont
  {Paarmann}}, \bibinfo {author} {\bibfnamefont {E.~S.}\ \bibnamefont
  {Zijlstra}}, \bibinfo {author} {\bibfnamefont {N.}~\bibnamefont {Karpowicz}},
  \bibinfo {author} {\bibfnamefont {M.~E.}\ \bibnamefont {Garcia}}, \bibinfo
  {author} {\bibfnamefont {R.}~\bibnamefont {Kienberger}}, \ and\ \bibinfo
  {author} {\bibfnamefont {R.}~\bibnamefont {Ernstorfer}},\ }\bibfield  {title}
  {\enquote {\bibinfo {title} {{Ultrafast Evolution of the Excited-State
  Potential Energy Surface of TiO$_2$ Single Crystals Induced by Carrier
  Cooling}},}\ }\href {\doibase 10.1103/PhysRevLett.110.067402} {\bibfield
  {journal} {\bibinfo  {journal} {Phys. Rev. Lett.}\ }\textbf {\bibinfo
  {volume} {110}},\ \bibinfo {pages} {067402} (\bibinfo {year}
  {2013})}\BibitemShut {NoStop}%
\bibitem [{\citenamefont {Giret}\ \emph {et~al.}(2011)\citenamefont {Giret},
  \citenamefont {Gell\'{e}},\ and\ \citenamefont {Arnaud}}]{Giret2011}%
  \BibitemOpen
  \bibfield  {author} {\bibinfo {author} {\bibfnamefont {Y.}~\bibnamefont
  {Giret}}, \bibinfo {author} {\bibfnamefont {A.}~\bibnamefont {Gell\'{e}}}, \
  and\ \bibinfo {author} {\bibfnamefont {B.}~\bibnamefont {Arnaud}},\
  }\bibfield  {title} {\enquote {\bibinfo {title} {{Entropy driven atomic
  motion in laser-excited bismuth}},}\ }\href {\doibase
  10.1103/PhysRevLett.106.155503} {\bibfield  {journal} {\bibinfo  {journal}
  {Physical Review Letters}\ }\textbf {\bibinfo {volume} {106}},\ \bibinfo
  {pages} {155503} (\bibinfo {year} {2011})}\BibitemShut {NoStop}%
\bibitem [{\citenamefont {Arnaud}\ and\ \citenamefont
  {Giret}(2013)}]{Arnaud2013}%
  \BibitemOpen
  \bibfield  {author} {\bibinfo {author} {\bibfnamefont {B.}~\bibnamefont
  {Arnaud}}\ and\ \bibinfo {author} {\bibfnamefont {Y.}~\bibnamefont {Giret}},\
  }\bibfield  {title} {\enquote {\bibinfo {title} {{Electron cooling and
  Debye-Waller effect in photoexcited bismuth}},}\ }\href {\doibase
  10.1103/PhysRevLett.110.016405} {\bibfield  {journal} {\bibinfo  {journal}
  {Physical Review Letters}\ }\textbf {\bibinfo {volume} {110}},\ \bibinfo
  {pages} {016405} (\bibinfo {year} {2013})}\BibitemShut {NoStop}%
\bibitem [{\citenamefont {Walker}(1956)}]{Walker1956}%
  \BibitemOpen
  \bibfield  {author} {\bibinfo {author} {\bibfnamefont {C.~B.}\ \bibnamefont
  {Walker}},\ }\bibfield  {title} {\enquote {\bibinfo {title} {{X-Ray Study of
  Lattice Vibrations}},}\ }\href {\doibase
  http://dx.doi.org/10.1103/PhysRev.103.547} {\bibfield  {journal} {\bibinfo
  {journal} {Physical Review}\ }\textbf {\bibinfo {volume} {103}},\ \bibinfo
  {pages} {547--557} (\bibinfo {year} {1956})}\BibitemShut {NoStop}%
\bibitem [{\citenamefont {Dacorogna}\ \emph {et~al.}(1985)\citenamefont
  {Dacorogna}, \citenamefont {Cohen},\ and\ \citenamefont
  {Lam}}]{Dacorogna1985}%
  \BibitemOpen
  \bibfield  {author} {\bibinfo {author} {\bibfnamefont {M.~M.}\ \bibnamefont
  {Dacorogna}}, \bibinfo {author} {\bibfnamefont {M.~L.}\ \bibnamefont
  {Cohen}}, \ and\ \bibinfo {author} {\bibfnamefont {P.~K.}\ \bibnamefont
  {Lam}},\ }\bibfield  {title} {\enquote {\bibinfo {title} {{Self-Consistent
  Calculation of the q Dependence of the Electron-Phonon Coupling in
  Aluminum}},}\ }\href {\doibase 10.1103/PhysRevLett.55.837} {\bibfield
  {journal} {\bibinfo  {journal} {Physical Review Letters}\ }\textbf {\bibinfo
  {volume} {55}},\ \bibinfo {pages} {837--840} (\bibinfo {year}
  {1985})}\BibitemShut {NoStop}%
\bibitem [{\citenamefont {Sears}\ and\ \citenamefont
  {Shelley}(1991)}]{Sears1991}%
  \BibitemOpen
  \bibfield  {author} {\bibinfo {author} {\bibfnamefont {V.~F.}\ \bibnamefont
  {Sears}}\ and\ \bibinfo {author} {\bibfnamefont {S.~A.}\ \bibnamefont
  {Shelley}},\ }\bibfield  {title} {\enquote {\bibinfo {title} {{Debye-Waller
  factor for elemental crystals}},}\ }\href {\doibase
  10.1107/S0108767391002970} {\bibfield  {journal} {\bibinfo  {journal} {Acta
  Crystallographica Section A}\ }\textbf {\bibinfo {volume} {47}},\ \bibinfo
  {pages} {441--446} (\bibinfo {year} {1991})}\BibitemShut {NoStop}%
\bibitem [{\citenamefont {Tang}\ \emph {et~al.}(2010)\citenamefont {Tang},
  \citenamefont {Li},\ and\ \citenamefont {Fultz}}]{Tang2010}%
  \BibitemOpen
  \bibfield  {author} {\bibinfo {author} {\bibfnamefont {X.}~\bibnamefont
  {Tang}}, \bibinfo {author} {\bibfnamefont {C.~W.}\ \bibnamefont {Li}}, \ and\
  \bibinfo {author} {\bibfnamefont {B.}~\bibnamefont {Fultz}},\ }\bibfield
  {title} {\enquote {\bibinfo {title} {{Anharmonicity-induced phonon broadening
  in aluminum at high temperatures}},}\ }\href {\doibase
  10.1103/PhysRevB.82.184301} {\bibfield  {journal} {\bibinfo  {journal}
  {Physical Review B}\ }\textbf {\bibinfo {volume} {82}},\ \bibinfo {pages}
  {184301} (\bibinfo {year} {2010})}\BibitemShut {NoStop}%
\bibitem [{\citenamefont {Beaurepaire}\ \emph {et~al.}(1996)\citenamefont
  {Beaurepaire}, \citenamefont {Merle}, \citenamefont {Daunois},\ and\
  \citenamefont {Bigot}}]{Beaurepaire1996}%
  \BibitemOpen
  \bibfield  {author} {\bibinfo {author} {\bibfnamefont {E.}~\bibnamefont
  {Beaurepaire}}, \bibinfo {author} {\bibfnamefont {J.-C.}\ \bibnamefont
  {Merle}}, \bibinfo {author} {\bibfnamefont {A.}~\bibnamefont {Daunois}}, \
  and\ \bibinfo {author} {\bibfnamefont {J.-Y.}\ \bibnamefont {Bigot}},\
  }\bibfield  {title} {\enquote {\bibinfo {title} {{Ultrafast Spin Dynamics in
  Ferromagnetic Nickel}},}\ }\href {\doibase 10.1103/PhysRevLett.76.4250}
  {\bibfield  {journal} {\bibinfo  {journal} {Physical Review Letters}\
  }\textbf {\bibinfo {volume} {76}},\ \bibinfo {pages} {4250} (\bibinfo {year}
  {1996})},\ \Eprint {http://arxiv.org/abs/9709264} {9709264} \BibitemShut
  {NoStop}%
\bibitem [{\citenamefont {Leguay}\ \emph {et~al.}(2013)\citenamefont {Leguay},
  \citenamefont {L\'{e}vy}, \citenamefont {Chimier}, \citenamefont
  {Deneuville}, \citenamefont {Descamps}, \citenamefont {Fourment},
  \citenamefont {Goyon}, \citenamefont {Hulin}, \citenamefont {Petit},
  \citenamefont {Peyrusse}, \citenamefont {Santos}, \citenamefont {Combis},
  \citenamefont {Holst}, \citenamefont {Recoules}, \citenamefont {Renaudin},
  \citenamefont {Videau},\ and\ \citenamefont {Dorchies}}]{Leguay2013}%
  \BibitemOpen
  \bibfield  {author} {\bibinfo {author} {\bibfnamefont {P.~M.}\ \bibnamefont
  {Leguay}}, \bibinfo {author} {\bibfnamefont {A.}~\bibnamefont {L\'{e}vy}},
  \bibinfo {author} {\bibfnamefont {B.}~\bibnamefont {Chimier}}, \bibinfo
  {author} {\bibfnamefont {F.}~\bibnamefont {Deneuville}}, \bibinfo {author}
  {\bibfnamefont {D.}~\bibnamefont {Descamps}}, \bibinfo {author}
  {\bibfnamefont {C.}~\bibnamefont {Fourment}}, \bibinfo {author}
  {\bibfnamefont {C.}~\bibnamefont {Goyon}}, \bibinfo {author} {\bibfnamefont
  {S.}~\bibnamefont {Hulin}}, \bibinfo {author} {\bibfnamefont
  {S.}~\bibnamefont {Petit}}, \bibinfo {author} {\bibfnamefont
  {O.}~\bibnamefont {Peyrusse}}, \bibinfo {author} {\bibfnamefont {J.~J.}\
  \bibnamefont {Santos}}, \bibinfo {author} {\bibfnamefont {P.}~\bibnamefont
  {Combis}}, \bibinfo {author} {\bibfnamefont {B.}~\bibnamefont {Holst}},
  \bibinfo {author} {\bibfnamefont {V.}~\bibnamefont {Recoules}}, \bibinfo
  {author} {\bibfnamefont {P.}~\bibnamefont {Renaudin}}, \bibinfo {author}
  {\bibfnamefont {L.}~\bibnamefont {Videau}}, \ and\ \bibinfo {author}
  {\bibfnamefont {F.}~\bibnamefont {Dorchies}},\ }\bibfield  {title} {\enquote
  {\bibinfo {title} {{Ultrafast Short-Range Disordering of
  Femtosecond-Laser-Heated Warm Dense Aluminum}},}\ }\href {\doibase
  10.1103/PhysRevLett.111.245004} {\bibfield  {journal} {\bibinfo  {journal}
  {Physical Review Letters}\ }\textbf {\bibinfo {volume} {111}},\ \bibinfo
  {pages} {245004} (\bibinfo {year} {2013})}\BibitemShut {NoStop}%
\bibitem [{\citenamefont {Allen}(1978)}]{Allen1978}%
  \BibitemOpen
  \bibfield  {author} {\bibinfo {author} {\bibfnamefont {P.~B.}\ \bibnamefont
  {Allen}},\ }\bibfield  {title} {\enquote {\bibinfo {title} {{New method for
  solving Boltzmann's equation for electrons in metals}},}\ }\href {\doibase
  10.1103/PhysRevB.17.3725} {\bibfield  {journal} {\bibinfo  {journal}
  {Physical Review B}\ }\textbf {\bibinfo {volume} {17}},\ \bibinfo {pages}
  {3725--3734} (\bibinfo {year} {1978})}\BibitemShut {NoStop}%
\bibitem [{\citenamefont {Chase}\ \emph {et~al.}(2016)\citenamefont {Chase},
  \citenamefont {Trigo}, \citenamefont {Reid}, \citenamefont {Li},
  \citenamefont {Vecchione}, \citenamefont {Shen}, \citenamefont {Weathersby},
  \citenamefont {Coffee}, \citenamefont {Hartmann}, \citenamefont {Reis},
  \citenamefont {Wang},\ and\ \citenamefont {D\"{u}rr}}]{Chase2016}%
  \BibitemOpen
  \bibfield  {author} {\bibinfo {author} {\bibfnamefont {T.}~\bibnamefont
  {Chase}}, \bibinfo {author} {\bibfnamefont {M.}~\bibnamefont {Trigo}},
  \bibinfo {author} {\bibfnamefont {A.~H.}\ \bibnamefont {Reid}}, \bibinfo
  {author} {\bibfnamefont {R.}~\bibnamefont {Li}}, \bibinfo {author}
  {\bibfnamefont {T.}~\bibnamefont {Vecchione}}, \bibinfo {author}
  {\bibfnamefont {X.}~\bibnamefont {Shen}}, \bibinfo {author} {\bibfnamefont
  {S.}~\bibnamefont {Weathersby}}, \bibinfo {author} {\bibfnamefont
  {R.}~\bibnamefont {Coffee}}, \bibinfo {author} {\bibfnamefont
  {N.}~\bibnamefont {Hartmann}}, \bibinfo {author} {\bibfnamefont {D.~A.}\
  \bibnamefont {Reis}}, \bibinfo {author} {\bibfnamefont {X.~J.}\ \bibnamefont
  {Wang}}, \ and\ \bibinfo {author} {\bibfnamefont {H.~A.}\ \bibnamefont
  {D\"{u}rr}},\ }\bibfield  {title} {\enquote {\bibinfo {title} {{Ultrafast
  electron diffraction from non-equilibrium phonons in femtosecond laser heated
  Au films}},}\ }\href {\doibase 10.1063/1.4940981} {\bibfield  {journal}
  {\bibinfo  {journal} {Applied Physics Letters}\ }\textbf {\bibinfo {volume}
  {108}},\ \bibinfo {pages} {041909} (\bibinfo {year} {2016})}\BibitemShut
  {NoStop}%
\bibitem [{\citenamefont {Trigo}\ \emph {et~al.}(2010)\citenamefont {Trigo},
  \citenamefont {Chen}, \citenamefont {Vishwanath}, \citenamefont {Sheu},
  \citenamefont {Graber}, \citenamefont {Henning},\ and\ \citenamefont
  {Reis}}]{Trigo2010}%
  \BibitemOpen
  \bibfield  {author} {\bibinfo {author} {\bibfnamefont {M.}~\bibnamefont
  {Trigo}}, \bibinfo {author} {\bibfnamefont {J.}~\bibnamefont {Chen}},
  \bibinfo {author} {\bibfnamefont {V.~H.}\ \bibnamefont {Vishwanath}},
  \bibinfo {author} {\bibfnamefont {Y.~M.}\ \bibnamefont {Sheu}}, \bibinfo
  {author} {\bibfnamefont {T.}~\bibnamefont {Graber}}, \bibinfo {author}
  {\bibfnamefont {R.}~\bibnamefont {Henning}}, \ and\ \bibinfo {author}
  {\bibfnamefont {D.~A.}\ \bibnamefont {Reis}},\ }\bibfield  {title} {\enquote
  {\bibinfo {title} {{Imaging nonequilibrium atomic vibrations with x-ray
  diffuse scattering}},}\ }\href {\doibase 10.1103/PhysRevB.82.235205}
  {\bibfield  {journal} {\bibinfo  {journal} {Physical Review B}\ }\textbf
  {\bibinfo {volume} {82}},\ \bibinfo {pages} {235205} (\bibinfo {year}
  {2010})},\ \Eprint {http://arxiv.org/abs/1006.3990} {1006.3990} \BibitemShut
  {NoStop}%
\bibitem [{abi()}]{abinit}%
  \BibitemOpen
  \href {http://www.abinit.org} {}\bibinfo {note}
  {Http://www.abinit.org}\BibitemShut {NoStop}%
\bibitem [{\citenamefont {Gonze}\ \emph {et~al.}(2009)\citenamefont {Gonze},
  \citenamefont {Amadon}, \citenamefont {Anglade}, \citenamefont {Beuken},
  \citenamefont {Bottin}, \citenamefont {Boulanger}, \citenamefont {Bruneval},
  \citenamefont {Caliste}, \citenamefont {Caracas}, \citenamefont
  {C\^{o}t\'{e}}, \citenamefont {Deutsch}, \citenamefont {Genovese},
  \citenamefont {Ghosez}, \citenamefont {Giantomassi}, \citenamefont
  {Goedecker}, \citenamefont {Hamann}, \citenamefont {Hermet}, \citenamefont
  {Jollet}, \citenamefont {Jomard}, \citenamefont {Leroux}, \citenamefont
  {Mancini}, \citenamefont {Mazevet}, \citenamefont {Oliveira}, \citenamefont
  {Onida}, \citenamefont {Pouillon}, \citenamefont {Rangel}, \citenamefont
  {Rignanese}, \citenamefont {Sangalli}, \citenamefont {Shaltaf}, \citenamefont
  {Torrent}, \citenamefont {Verstraete}, \citenamefont {Zerah},\ and\
  \citenamefont {Zwanziger}}]{Gonze:2009}%
  \BibitemOpen
  \bibfield  {author} {\bibinfo {author} {\bibfnamefont {X.}~\bibnamefont
  {Gonze}}, \bibinfo {author} {\bibfnamefont {B.}~\bibnamefont {Amadon}},
  \bibinfo {author} {\bibfnamefont {P.-M.}\ \bibnamefont {Anglade}}, \bibinfo
  {author} {\bibfnamefont {J.-M.}\ \bibnamefont {Beuken}}, \bibinfo {author}
  {\bibfnamefont {F.}~\bibnamefont {Bottin}}, \bibinfo {author} {\bibfnamefont
  {P.}~\bibnamefont {Boulanger}}, \bibinfo {author} {\bibfnamefont
  {F.}~\bibnamefont {Bruneval}}, \bibinfo {author} {\bibfnamefont
  {D.}~\bibnamefont {Caliste}}, \bibinfo {author} {\bibfnamefont
  {R.}~\bibnamefont {Caracas}}, \bibinfo {author} {\bibfnamefont
  {M.}~\bibnamefont {C\^{o}t\'{e}}}, \bibinfo {author} {\bibfnamefont
  {T.}~\bibnamefont {Deutsch}}, \bibinfo {author} {\bibfnamefont
  {L.}~\bibnamefont {Genovese}}, \bibinfo {author} {\bibfnamefont {Ph.}\
  \bibnamefont {Ghosez}}, \bibinfo {author} {\bibfnamefont {M.}~\bibnamefont
  {Giantomassi}}, \bibinfo {author} {\bibfnamefont {S.}~\bibnamefont
  {Goedecker}}, \bibinfo {author} {\bibfnamefont {D.~R.}\ \bibnamefont
  {Hamann}}, \bibinfo {author} {\bibfnamefont {P.}~\bibnamefont {Hermet}},
  \bibinfo {author} {\bibfnamefont {F.}~\bibnamefont {Jollet}}, \bibinfo
  {author} {\bibfnamefont {G.}~\bibnamefont {Jomard}}, \bibinfo {author}
  {\bibfnamefont {S.}~\bibnamefont {Leroux}}, \bibinfo {author} {\bibfnamefont
  {M.}~\bibnamefont {Mancini}}, \bibinfo {author} {\bibfnamefont
  {S.}~\bibnamefont {Mazevet}}, \bibinfo {author} {\bibfnamefont {M.~J.~T.}\
  \bibnamefont {Oliveira}}, \bibinfo {author} {\bibfnamefont {G.}~\bibnamefont
  {Onida}}, \bibinfo {author} {\bibfnamefont {Y.}~\bibnamefont {Pouillon}},
  \bibinfo {author} {\bibfnamefont {T.}~\bibnamefont {Rangel}}, \bibinfo
  {author} {\bibfnamefont {G.-M.}\ \bibnamefont {Rignanese}}, \bibinfo {author}
  {\bibfnamefont {D.}~\bibnamefont {Sangalli}}, \bibinfo {author}
  {\bibfnamefont {R.}~\bibnamefont {Shaltaf}}, \bibinfo {author} {\bibfnamefont
  {M.}~\bibnamefont {Torrent}}, \bibinfo {author} {\bibfnamefont {M.J.}\
  \bibnamefont {Verstraete}}, \bibinfo {author} {\bibfnamefont
  {G.}~\bibnamefont {Zerah}}, \ and\ \bibinfo {author} {\bibfnamefont {J.W.}\
  \bibnamefont {Zwanziger}},\ }\bibfield  {title} {\enquote {\bibinfo {title}
  {Abinit: First-principles approach to material and nanosystem properties},}\
  }\href {\doibase doi:10.1016/j.cpc.2009.07.007} {\bibfield  {journal}
  {\bibinfo  {journal} {Computer Physics Communications}\ }\textbf {\bibinfo
  {volume} {180}},\ \bibinfo {pages} {2582} (\bibinfo {year}
  {2009})}\BibitemShut {NoStop}%
\bibitem [{\citenamefont {Gonze}(1997)}]{Gonze:1997a}%
  \BibitemOpen
  \bibfield  {author} {\bibinfo {author} {\bibfnamefont {X.}~\bibnamefont
  {Gonze}},\ }\bibfield  {title} {\enquote {\bibinfo {title} {First-principles
  responses of solids to atomic displacements and homogeneous electric fields:
  implementation of a conjugate-gradient algorithm},}\ }\href {\doibase
  10.1103/PhysRevB.55.10337} {\bibfield  {journal} {\bibinfo  {journal} {Phys.
  Rev. B}\ }\textbf {\bibinfo {volume} {55}},\ \bibinfo {pages} {10337}
  (\bibinfo {year} {1997})}\BibitemShut {NoStop}%
\bibitem [{\citenamefont {Gonze}\ and\ \citenamefont
  {Lee}(1997)}]{Gonze:1997b}%
  \BibitemOpen
  \bibfield  {author} {\bibinfo {author} {\bibfnamefont {X.}~\bibnamefont
  {Gonze}}\ and\ \bibinfo {author} {\bibfnamefont {C.}~\bibnamefont {Lee}},\
  }\bibfield  {title} {\enquote {\bibinfo {title} {Dynamical matrices, born
  effective charges, dielectric permittivity tensors, and interatomic force
  constants from density-functional perturbation theory},}\ }\href {\doibase
  10.1103/PhysRevB.55.10355} {\bibfield  {journal} {\bibinfo  {journal} {Phys.
  Rev. B}\ }\textbf {\bibinfo {volume} {55}},\ \bibinfo {pages} {10355}
  (\bibinfo {year} {1997})}\BibitemShut {NoStop}%
\bibitem [{\citenamefont {Savrasov}\ and\ \citenamefont
  {Savrasov}(1996)}]{Savrasov:1996}%
  \BibitemOpen
  \bibfield  {author} {\bibinfo {author} {\bibfnamefont {S.~Y.}\ \bibnamefont
  {Savrasov}}\ and\ \bibinfo {author} {\bibfnamefont {D.~Y.}\ \bibnamefont
  {Savrasov}},\ }\bibfield  {title} {\enquote {\bibinfo {title}
  {Electron-phonon interactions and related physical properties of metals from
  linear-response theory},}\ }\href {\doibase
  http://dx.doi.org/10.1103/PhysRevB.54.16487} {\bibfield  {journal} {\bibinfo
  {journal} {Phys. Rev. B}\ }\textbf {\bibinfo {volume} {54}},\ \bibinfo
  {pages} {16487} (\bibinfo {year} {1996})}\BibitemShut {NoStop}%
\bibitem [{\citenamefont {Fuchs}\ and\ \citenamefont
  {Scheffler}(1999)}]{Fuchs:1999}%
  \BibitemOpen
  \bibfield  {author} {\bibinfo {author} {\bibfnamefont {M.}~\bibnamefont
  {Fuchs}}\ and\ \bibinfo {author} {\bibfnamefont {M.}~\bibnamefont
  {Scheffler}},\ }\bibfield  {title} {\enquote {\bibinfo {title} {Ab initio
  pseudopotentials for electronic structure calculations of poly-atomic systems
  using density-functional theory},}\ }\href {\doibase
  doi:10.1016/S0010-4655(98)00201-X} {\bibfield  {journal} {\bibinfo  {journal}
  {Comput. Phys. Commun.}\ }\textbf {\bibinfo {volume} {119}},\ \bibinfo
  {pages} {67} (\bibinfo {year} {1999})}\BibitemShut {NoStop}%
\bibitem [{\citenamefont {Perdew}\ \emph {et~al.}(1996)\citenamefont {Perdew},
  \citenamefont {Burke},\ and\ \citenamefont {Ernzerhof}}]{perdew:1996}%
  \BibitemOpen
  \bibfield  {author} {\bibinfo {author} {\bibfnamefont {J~P.}\ \bibnamefont
  {Perdew}}, \bibinfo {author} {\bibfnamefont {K}~\bibnamefont {Burke}}, \ and\
  \bibinfo {author} {\bibfnamefont {M}~\bibnamefont {Ernzerhof}},\ }\bibfield
  {title} {\enquote {\bibinfo {title} {Generalized gradient approximation made
  simple},}\ }\href {\doibase http://dx.doi.org/10.1103/PhysRevLett.77.3865}
  {\bibfield  {journal} {\bibinfo  {journal} {Phys. Rev. Lett.}\ }\textbf
  {\bibinfo {volume} {77}},\ \bibinfo {pages} {3865} (\bibinfo {year}
  {1996})}\BibitemShut {NoStop}%
\bibitem [{\citenamefont {Smr{\v c}ka}(1970)}]{Smrcka1970}%
  \BibitemOpen
  \bibfield  {author} {\bibinfo {author} {\bibfnamefont {L.}~\bibnamefont
  {Smr{\v c}ka}},\ }\bibfield  {title} {\enquote {\bibinfo {title} {{Energy
  band structure of aluminium by the augmented plane wave method}},}\
  }\href@noop {} {\bibfield  {journal} {\bibinfo  {journal} {Czech. J. Phys.
  B}\ }\textbf {\bibinfo {volume} {20}},\ \bibinfo {pages} {291--300} (\bibinfo
  {year} {1970})}\BibitemShut {NoStop}%
\bibitem [{\citenamefont {Levinson}\ \emph {et~al.}(1983)\citenamefont
  {Levinson}, \citenamefont {Greuter},\ and\ \citenamefont
  {Plummer}}]{Levinson1983}%
  \BibitemOpen
  \bibfield  {author} {\bibinfo {author} {\bibfnamefont {H.~J.}\ \bibnamefont
  {Levinson}}, \bibinfo {author} {\bibfnamefont {F.}~\bibnamefont {Greuter}}, \
  and\ \bibinfo {author} {\bibfnamefont {E.~W.}\ \bibnamefont {Plummer}},\
  }\bibfield  {title} {\enquote {\bibinfo {title} {{Experimental band structure
  of aluminum}},}\ }\href {\doibase 10.1103/PhysRevB.27.727} {\bibfield
  {journal} {\bibinfo  {journal} {Physical Review B}\ }\textbf {\bibinfo
  {volume} {27}},\ \bibinfo {pages} {727--747} (\bibinfo {year}
  {1983})}\BibitemShut {NoStop}%
\bibitem [{\citenamefont {Nechaev}\ \emph {et~al.}(2008)\citenamefont
  {Nechaev}, \citenamefont {Sklyadneva}, \citenamefont {Silkin}, \citenamefont
  {Echenique},\ and\ \citenamefont {Chulkov}}]{Nechaev2008}%
  \BibitemOpen
  \bibfield  {author} {\bibinfo {author} {\bibfnamefont {I.~A.}\ \bibnamefont
  {Nechaev}}, \bibinfo {author} {\bibfnamefont {I.~Yu.}\ \bibnamefont
  {Sklyadneva}}, \bibinfo {author} {\bibfnamefont {V.~M.}\ \bibnamefont
  {Silkin}}, \bibinfo {author} {\bibfnamefont {P.~M.}\ \bibnamefont
  {Echenique}}, \ and\ \bibinfo {author} {\bibfnamefont {E.~V.}\ \bibnamefont
  {Chulkov}},\ }\bibfield  {title} {\enquote {\bibinfo {title} {{Theoretical
  study of quasiparticle inelastic lifetimes as applied to aluminum}},}\ }\href
  {\doibase 10.1103/PhysRevB.78.085113} {\bibfield  {journal} {\bibinfo
  {journal} {Physical Review B}\ }\textbf {\bibinfo {volume} {78}},\ \bibinfo
  {pages} {085113} (\bibinfo {year} {2008})}\BibitemShut {NoStop}%
\bibitem [{\citenamefont {McMillan}(1968)}]{McMillan:1968}%
  \BibitemOpen
  \bibfield  {author} {\bibinfo {author} {\bibfnamefont {W.~L.}\ \bibnamefont
  {McMillan}},\ }\bibfield  {title} {\enquote {\bibinfo {title} {Transition
  temperature of strong-coupled superconductors},}\ }\href {\doibase
  10.1103/PhysRev.167.331} {\bibfield  {journal} {\bibinfo  {journal} {Phys.
  Rev.}\ }\textbf {\bibinfo {volume} {167}},\ \bibinfo {pages} {331--344}
  (\bibinfo {year} {1968})}\BibitemShut {NoStop}%
\end{thebibliography}%
\end{document}